\documentclass[aps,prd,preprintnumbers,showpacs,nofootinbib,amssymb]{revtex4}

\usepackage{amsfonts,amsmath,units,wasysym,epsfig,graphicx,verbatim,color,subfigure,graphicx}
\usepackage{amsmath}
\usepackage{amssymb}
\usepackage{amsfonts}
\usepackage{bm}
\usepackage{color}

\def\be{\begin{equation}}
\def\ee{\end{equation}}
\def\bea{\begin{eqnarray}}
\def\eea{\end{eqnarray}}

\def \f0{f_{0}}
\def \t0{\tau_{0}}
\def \tautemp{\tau_{\rm template}}
\def \M{{\cal M}}
\def \Msun{M_{\odot}}
\def \fl{f_{\rm lower}}
\def \fu{f_{\rm upper}}
\def \fs{f_{s}}
\def \h {\frac{1}{2}}
\def \ssg{\sigma_{sg}}
\def \sf{\sigma_{\phi}}

\def\lsim{\mathrel{\rlap{\lower4pt\hbox{\hskip1pt$\sim$}}
    \raise1pt\hbox{$<$}}}                
\def\gsim{\mathrel{\rlap{\lower4pt\hbox{\hskip1pt$\sim$}}
    \raise1pt\hbox{$>$}}}                

\begin{document}

\newcommand{\rhat}{\hat{r}}
\newcommand{\iotahat}{\hat{\iota}}
\newcommand{\phihat}{\hat{\phi}}
\newcommand{\IUCAA}{Inter-University Centre for Astronomy and
  Astrophysics, Post Bag 4, Ganeshkhind, Pune 411 007, India}
\newcommand{\IUCAAB}{Inter-University Centre for Astronomy and Astrophysics, Post Bag 4, Ganeshkhind, Pune 411 007, India}

\newcommand{\WSU}{Department of Physics \& Astronomy, Washington State University,
1245 Webster, Pullman, WA 99164-2814, U.S.A \\}
\newcommand{\AEI}{Max-Planck-Institut f\"{u}r Gravitationsphysik,
  Callinstrasse 38, D-30167 Hannover, Germany}

\title{Towards mitigating the effect of sine-Gaussian noise transients on searches for gravitational waves from compact binary coalescences}

\author{Sukanta Bose${}^{1,2}$, Sanjeev Dhurandhar${}^{1}$, Anuradha
  Gupta${}^{1}$, Andrew Lundgren${}^{3}$}
\affiliation{\IUCAA}
\affiliation{\WSU}
\affiliation{\AEI}

\date{\today}


\begin{abstract}

Gravitational wave (GW) signals were recently detected directly by
LIGO from the coalescences of two stellar mass black hole pairs. These
detections have strengthened our long held belief that compact binary
coalescences (CBCs) are the most promising GW detection prospects
accessible to ground-based interferometric detectors. 
For detecting CBC signals it is of vital importance to characterize
and identify non-Gaussian and non-stationary noise in these
detectors. In this work we model two important classes
of transient artifacts that contribute to this noise and adversely
affect the detector sensitivity to CBC signals. One of them is the
sine-Gaussian “glitch”, characterized by a central frequency $f_0$ and
a quality factor $Q$ and the other is the chirping sine-Gaussian
glitch, which is characterized by $f_0$, $Q$ as well as a chirp
parameter. We study the response a bank of compact binary inspiral
templates has to these two families of glitches when they are used to
match-filter data containing any of these glitches. Two important
characteristics of this response are the distributions of
the signal-to-noise ratio (SNR) and the time-lag (i.e., how long after
the occurrence of a glitch the SNR of a trigger arising from its
matched-filtering by a template peaks) of individual
templates. We show how these distributions differ from those when the
detector data has a real CBC signal instead of a glitch. We argue that
these distinctions can be utilized to develop useful signal-artifact
discriminators that add negligibly to the computational cost of a CBC
search. 
Specifically, we show how the central frequency of a glitch
can be used to set adaptive time-windows around it so that any
template trigger occurring in that window can be quarantined for
further vetting of its supposed astrophysical nature. Second, we
recommend focusing efforts on reducing the incidence of glitches with
low central-frequency values because they create CBC triggers with the
longest time-lags. This work allows us to associate such triggers with the glitches 
which otherwise would have escaped attention.

\end{abstract}

\preprint{[LIGO-P1600145]}

\maketitle

\section{Introduction}

The two detectors of the Advanced Laser Interferometer Gravitational
Wave Observatory (aLIGO)~\cite{TheLIGOScientific:2014jea},
located in Hanford (H1), Washington, and
Livingston (L1), Louisiana, completed their first observation run
``O1'' in January 2016. It had begun in September 2015, and lasted 
for about four months. A gravitational wave (GW) signal from the coalescence of two 
black holes, of about 30 $M_\odot$ each, was observed by the two
detectors on September 14, 2015, during the eighth engineering run,
barely a few days before the commencement of O1. This is not only the first
direct detection of GWs but also the observation of the heaviest stellar mass
black holes to date and the first discovery of a binary black hole
merger \cite{DetGW}. A GW signal from a second binary black hole coalescence was
detected by LIGO on December 26, 2015~\cite{GW2}, thereby, firmly
launching a new era in astronomy.
\par

\begin{figure}[h!]
\begin{center}
\includegraphics[width=3.5in]{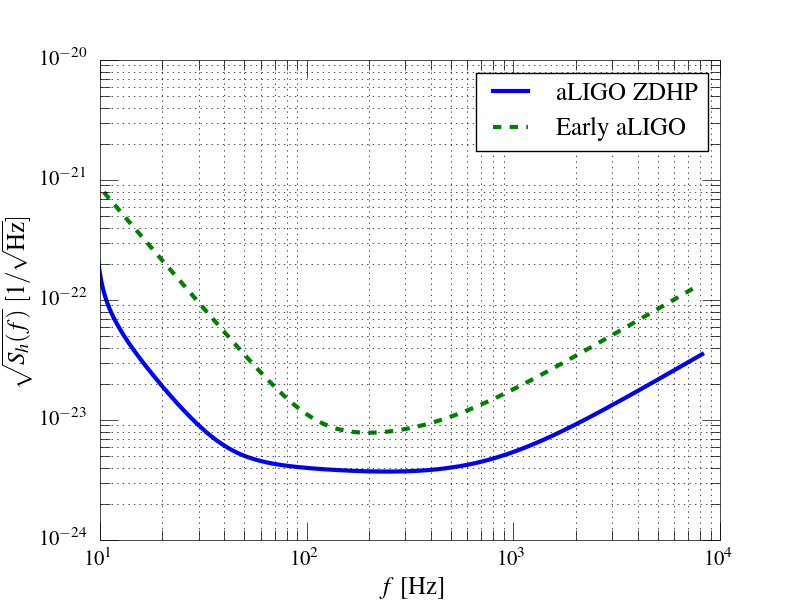}
\end{center}
\caption{Design sensitivities of aLIGO detectors. The dashed
  (green) line is the amplitude spectral density of early aLIGO,
  whereas the solid (blue) line is the same
  quantity plotted for the ``zero-detuned, high power'' (ZDHP)
  configuration of aLIGO, as a function of frequency. Here,
  $S_{h}(f)$ is the noise power-spectral density.
} 
\label{fig:noisepsds}
\end{figure}

The detectors are yet to reach their 
peak sensitivity, which will be achieved in the next few years. 
That will lay the foundation for what should eventually become a
ten-fold improvement in strain sensitivity and a thousand-fold
improvement in the event rate, relative to Initial LIGO. In this way, aLIGO is 
expected to establish the era of GW astronomy with multiple observations. 
Predictions of GW detection rates and estimation of source
localization errors with aLIGO, especially associated with compact
binary coalescences (CBCs)~\cite{Ratespaper},
are made based on a spectral
profile for the detector noise, such as the ones given in
Fig. \ref{fig:noisepsds}, and
the assumption that the noise is Gaussian and stationary. In reality
it is neither of those two (see, e.g., Ref.~\cite{Aasi:2014mqd}),
and can cause the predictions to come up short. This is why
the sources of non-Gaussianity and non-stationarity must be detected
and eliminated (see, e.g., Ref.~\cite{Bose:2016sqv} and the references therein). This activity will be critical in ensuring good observational depth
and thereby help toward realizing the expected rate of GW
detections. As commissioners work on lowering the noise floor in order
to approach that sensitivity, new types of broad and narrow band
noise features will come to the fore that will require identification,
analysis, modeling and eventual regression or mitigation. 
\par

As confirmed by the recent detections of GW events, the most promising source  
of GWs for ground-based interferometric detectors is the coalescence of the 
compact binary consisting of neutron stars and/or black holes. The
attractiveness of CBCs stems from the fact
that their signals are theoretically well modeled 
and their expected rates in the range of aLIGO are reasonably high~\cite{Ratespaper}.
\par
One of the primary types of noise that adversely affects the
sensitivity of searches for GWs from CBCs is a transient burst that is
either instrumental or environmental in origin and can be modeled as a
sine-Gaussian with a specific central frequency $f_0$ and quality
factor $Q$, or a combination of sine-Gaussians with different values
of $f_0$ and $Q$. These frequencies and quality factors are ultimately related
to the resonances and damping time-scales, respectively, of their sources.
For the same reason they also show up on Omega scans, which plots time-frequency maps of 
excess noise transients by resolving them in a sine-Gaussian basis \cite{Chatterji:2005thesis,Chatterji:2004qg}.
To search large amounts of data using sine-Gaussian basis, we use a reimplementation of the Omega algorithm known as Omicron. This is the main reason why we extensively study and compare the 
effect of sine-Gaussian and related glitches on CBC searches here. 

A theoretical framework was developed in Ref.~\cite{Canton:2013joa} to study
the effect of a ``sine-Gaussian'' strain 
on a non-spinning Newtonian inspiral template.
Here we extend that work in the following ways.
First, we study how a bank of thousands of non-spinning CBC templates
responds to a sine-Gaussian glitch injected into data sets when
they are analysed by a search pipeline that was commissioned for
aLIGO and Advanced Virgo (AdV) searches~\cite{TheVirgo:2014hva}. 
This pipeline is called PyCBC~\cite{Canton:2014ena}
and it uses matched filtering \cite{Sathyaprakash:1991mt}
with theoretically modeled CBC templates to search for CBC signals. 
When the signal-to-noise ratio (SNR) of the matched-filter output is
above a threshold value, the pipeline records the time of its occurrence and
other related parameters, such as the CBC masses of the template, as a
{\em trigger}.
Second, we repeat the above study for chirping sine-Gaussians, which
are similar to sine-Gaussians except that the phase of the sine
function now increases quadratically, instead of linearly, with time.
These types of noise transients can be hazardous to an astrophysical 
search because they can produce high SNR triggers 
if their frequency and the chirping rate match with those of the
template, as we show below. Third, we extend  the theoretical
framework that was developed in Ref.~\cite{Canton:2013joa} to
explain a CBC template-bank's response to both kinds of glitches. In doing
so, we qualitatively show how it can be used to
discriminate between such glitches and real CBC signals. We argue that
such discrimination can be made with negligible increase in
computational costs, which is helpful in light of the fact 
that GW trigger alerts need to be communicated to electromagnetic and
particle observatories for quick follow-ups for possible detection of
prompt emissions and afterglows~\cite{Rana:2016crg,Ghosh:2013yda}.
We also incorporate higher order post-Newtonian (PN) terms in the template to 
discern their effects on the response. 

Omicron automatically runs on multiple detector channels, including 
the GW channel, as data is acquired during an engineering or observation run, to 
identify noise glitches in them and archive them in a database. Data 
quality scripts, such as the Hierarchical Veto \cite{Smith:2011an}, or
``H-veto'' run on 
these triggers to flag times in the GW data when its quality might be 
suspect, e.g., due to influences from the environmental, instrumental 
or auxiliary channels. H-veto looks for coincidences of 
excess noise events (such as the triggers produced by Omicron) in the 
GW channel with those in the latter channels. These flags are used by
data analysis tools that search for astrophysical signals, e.g., from
CBCs~\cite{Usman:2015kfa,Messick:2016aqy,Allen:2004gu}, to interpret the GW triggers arising from glitchy data-sets with
greater care. 
Our study on trigger time-lags and SNRs will help computationally inexpensive 
tools like H-veto to set the time-windows for checking 
trigger coincidences in a more informed way. 
\par
The layout of the paper is as follows. In Sec. \ref{sec:sgbursts}, we recapitulate
the main results from the theoretical framework that was developed in Ref.~\cite{Canton:2013joa}.
In the process, we introduce notation that we follow in the rest of
the paper. Additionally, we present results from an astrophysical search 
running on strain data and explain them in terms of the framework. In
Sec. \ref{sec:csgbursts}, we present a similar study for chirping
sine-Gaussians, compare the results with non-chirping sine-Gaussians,
and analytically explain the reason behind the differences in the
responses of the same CBC template bank to these two classes of glitches. In
Sec. \ref{sec:vetoes}, we describe how our results can be applied to the
construction of computationally inexpensive 
tests for
discriminating real signals from noise glitches. Finally, we summarize the results in Sec. \ref{sec:conclusions}.

\section{Sine-Gaussian glitches and their effect on Inspiral
  Templates}
\label{sec:sgbursts}

Past experience with real data \cite{Aasi:2014mqd} 
has shown that one of the types of noise artifacts that affects the
background of the CBC searches is a
temporally localized sinusoidal burst that can be characterized by a frequency, $f_0$ and a
time-scale, which gives the duration of the rise and fall of its
amplitude. (See, e.g., Refs.~\cite{Mukherjee:2012bua,Powell:2015ona} and the
references therein.) 
There are multiple causes behind these transients, such as mechanical resonances, servo oscillations, etc. 
The time-scale of these transients is often alternatively described in
terms of a quality factor $Q$, such that for a fixed $f_0$ large
values of $Q$ correspond to large durations of the noise transient. 
These two parameters also define a sine-Gaussian
function. Indeed, as was shown in
Refs.~\cite{Chatterji:2005thesis,Chatterji:2004qg}, the amplitude and
phase of the detector strain corresponding to such noise
artifacts, or ``glitches'', can be expanded in a basis of
sine-Gaussians, of varying $f_0$ and $Q$. The third parameter that
characterizes these glitches is their time of occurrence.

\begin{figure}[h!]
\begin{center}
\includegraphics[width=3.5in]{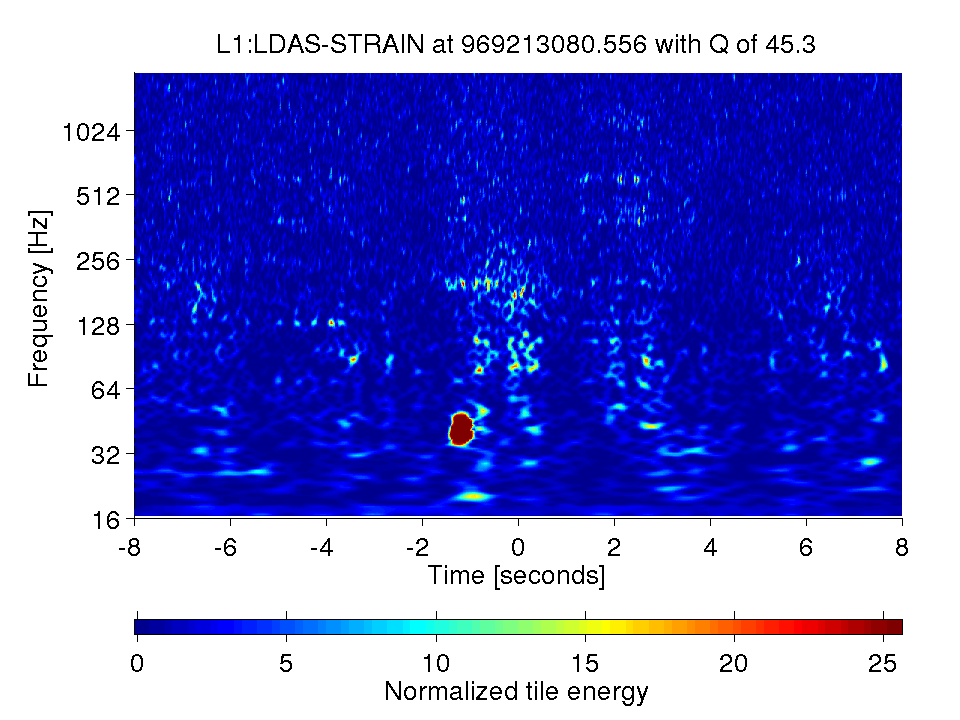}
\end{center}
\caption{The spectrogram shows a noise transient, located at
  approximately -1.2 sec in the time coordinate used above, that 
appeared in the gravitational-wave channel of the LIGO
  detector in Livingston (L1) during its sixth science run (S6). This
  figure was made by Omega scan and shows that the loudest sine-Gaussian
 component of this transient has $f_0 = 39.3$ Hz and $Q=45.3$. 
}
\label{fig:omicronSpect}
\end{figure}

In this section we extend the work initiated in
Ref.~\cite{Canton:2013joa} to study of the effect of sine-Gaussian
glitches in interferometric data on CBC searches. 
Such burst like noise transients are frequently seen in spectrograms 
created by tools like Omicron, as shown in Fig.~\ref{fig:omicronSpect}. 
The loudest sine-Gaussian component in this glitch was found to have
$f_0 =39.3$ Hz and $Q = 45.3$. The effect of the same sine-Gaussian 
on CBC templates, as studied by running the CBC search pipeline on it,
is shown in the right hand panel of Fig. \ref{fig:simSNRVsTimelag4sgburst}, 
which plots the SNRs of the templates triggered by the glitch
as a function of the time difference between the glitch and the end
time of the trigger, time-lag. Owing to the large SNRs produced and the multiple
templates triggered, these glitches create multiple false-alarms and
demand additional computation for identifying them and vetoing them.
Therefore, it is desirable to investigate and understand how these
sine-Gaussian glitches affect searches of CBC signals, even when the
templates employed use only the inspiral phase, e.g., in binary
neutron star searches, which have insignificant power in the merger
and ringdown part of the signal. 
\par
A CBC search is performed over the binary parameter space by employing
the matched filtering technique. This space comprises primarily two
functions of the binary's component masses for cases where the components are non-spinning. The effect of a
sine-Gaussian glitch is seen in the matched filter output when it is
matched filtered with a CBC template. Since the binary parameters are
not known {\it a priori}, a bank of multiple templates is used to
adequately cover the required parameter space. We consider nonspinning Newtonian amplitude waveforms.
In this case the waveform depends only on a particular combination of
individual masses of the binary - the chirp mass $\M$ - in addition to other kinematical  parameters. 
The sine-Gaussian is characterized by its central frequency $\f0$ and decay time $\tau$. 
Instead of $\tau$ we prefer the quality factor $Q$ as a parameter in our analysis. 
\par
In Ref.~\cite{Canton:2013joa} we had obtained the matched filter output as a function of time, denoted by $C(t)$, which is a complex function of the time-lag $t$, between the signal and the template. The quantity we were interested 
in was the modulus of $C(t)$ because then the maximization over the initial phase of the compact binary signal is automatically taken care of. 
Then we had compared it with analytical approximations in various regions 
of the parameter space. Here we are interested in the trigger generated by the matched filter - the maximum of 
$\vert C(t) \vert$ - when it encounters a sine-Gaussian glitch. 
The purpose of this section is to present analytical formulae for
those characteristics of a trigger that can be used to discriminate
the glitch from a CBC signal, namely, the SNR and the time-lag. 

We consider a sine-Gaussian glitch characterized by central frequency $\f0$ and quality factor $Q$:
\begin{align}
\label{eq:sg}
s(t)=  s_0 e^{-\frac{t^{2}}{ \tau^{2}}} \sin 2 \pi \f0 t \,,
\end{align}
where $s_0$ is the amplitude (which is set to unity, unless otherwise
specified), $\tau$ is the decay time-constant, which is related to the 
quality factor $Q$ by the relation 
$Q = 2 \pi f_0 \tau$. \footnote{Note that the $Q$ in this paper is
  $\sqrt{2}$ times the $Q$ defined by Ref.~\cite{Saurav:2004}.}
In reality, the glitches can come with a variety of amplitudes.  
Such a glitch produces a trail of triggers, especially, if it is a
strong one. This is because it gives significant cross-correlation
with a sizable subset of a CBC template bank owing to its power. This
effect is seen in the right hand panel in Fig. \ref{fig:simSNRVsTimelag4sgburst}.
\par
In the positive frequency domain $s(t)$ translates to a Gaussian around $\f0$, given by 
\begin{align}
\label{Eq_SG_freq}
{\tilde s}(f)= \left (\frac{1}{4 i \sqrt {\pi}} \frac{Q}{\f0} \right ) e^{- \frac{(f-\f0)^{2} Q^{2}}{4 \f0^{2}}} \,.
\end{align}
\par
We use a normalized template corresponding to a Newtonian binary
inspiral waveform for our analysis: 
\begin{align}
\label{eq:newtonianinspiral}
{\tilde h}(f)=h_{0} f^{-\frac{7}{6}} e^{-i \psi (f)} \,,
\end{align}
where $h_{0}$ is the normalization constant determined through the normalization equation,
\begin{align}
\label{eq:normalization}
\int_{\fl}^{\fu}  \frac{|{\tilde h}(f)|^2}{S_{h}(f)} \,df = 1 \,,
\end{align}
and $S_h(f)$ is the power spectral density (PSD) of the noise. In this
paper, we mainly use the zero-detuned, high power (ZDHP) PSD of 
aLIGO~\cite{aLIGOZDHP}, if not specifically stated otherwise.
The phase $\psi(f)$ is given by (see, e.g., Ref.~\cite{Pai:2000zt}):
\begin{align}
\label{eq:phase}
\psi(f)=2 \pi f t_{c} - \phi_{c} - \frac{\pi}{4} + \frac{3}{128}(\pi \M f)^{-\frac{5}{3}} \,,
\end{align}
where $\M$ is the chirp mass of the binary system, and  $t_c$ and $\phi_c$ are the coalescence time and phase, respectively. If $m_1$ and $m_2$
are the binary component masses, then $\M =(m_1m_2)^{3/5}/(m_1+m_2)^{1/5}$.
We henceforth set $t_{c}$ and $\phi_{c}$ to zero so that the chirp appears at the end of the time-series. 
\par
Further, we express the inspiral binary template in terms 
the chirp time $\t0$, as defined in Refs.~\cite{Sathyaprakash:1991mt,
  Sathyaprakash:1994}, instead of the chirp mass $\M$. As shown below,
$\t0$ will turn out to be a suitable unit for expressing the time-lag
between the occurrence of a glitch and the time of the trigger arising from
its cross-correlation with that template. The chirp time is used to
construct template banks because unit-norm templates
with a fixed minimal match (which, e.g., can be set to 97\%) with
neighboring templates are uniformly spaced in that parameter.
Physically, $\t0$ is the time taken for the binary to coalesce, starting from some fiducial 
frequency $f_{a}$. Here, the natural choice for $f_a$ is the central frequency of the sine-Gaussian, $\f0$. 
The chirp time $\t0$ with $\f0$ as the choice of fiducial frequency is defined as follows:
\be
\t0 = \frac{5}{256 \pi f_0} (\pi \M f_0)^{-5/3} \simeq 11.72 \left(
  \frac{f_{0}}{60~{\rm Hz}} \right)^{-8/3} \left( \frac{\M}{\Msun} \right)^{-5/3} {\rm sec} \,.
\ee
In terms of $\t0$, the phase of the inspiral template, given in Eq.~(\ref{eq:phase}), becomes
\begin{align}
\psi(f)= \frac{6 \pi \f0 \t0}{5} \left( \frac{f}{\f0} \right)^{-5/3} - \frac{\pi}{4} \,.
\end{align} 
The ranges of $f_0$, chirp mass, and quality factor that we use in our
study are $30$ Hz $\lesssim \f0 \lesssim 200$ Hz,  $0.87 M_{\odot}
\lesssim \M \lesssim  3M_{\odot}$ and
$5 \lesssim Q \lesssim 50$, respectively, unless otherwise
specified. They were arrived upon after a more expansive study that
went beyond these ranges. The final choices were made based partly on 
glitch parameter values observed in real data and partly on where in
parameter spaces of the signals and the glitches the longest
time-lags are found. 

\subsection{The correlation statistic}
\label{Sec_SG_match}

We briefly review the main results of Ref.~\cite{Canton:2013joa},
especially, those that are required for our analysis here. 
The matched filtering operation results in the complex correlation $C(t)$, where $t$ denotes the time-lag parameter between the signal and the template. The correlation $C(t)$ is given by:
\be
\label{crosscorr}
C(t) = \int_{\fl}^{\fu} \frac{{\tilde s}(f) {\tilde h}^{*}(f)}{S_{h}(f)} e^{2 \pi i f t} \,df 
     = \frac{Q h_{0} e^{-i \pi/4}}{4 i \sqrt{\pi} \f0} \int_{\fl}^{\fu} \frac{e^{- \frac{(f-\f0)^{2} Q^{2}}{4 \f0^{2}}} e^{i \phi(f)}}{f^{7/6} S_{h}(f)}~df \,.
\ee
The range of integration is from $\fl$ to $\fu$ with $\fl$ equal to the seismic cut-off frequency of aLIGO, i.e., 10~Hz and $\fu$ is equal to the frequency corresponding 
to the inner most stable circular orbit. The phase function $\phi$ is
\begin{align}
\phi(f)=2 \pi f t + \frac{6 \pi \f0 \t0}{5} \left(\frac{f}{\f0}\right)^{-5/3} \,.
\label{phase}
\end{align}
We Taylor expand the phase term $\phi(f)$ up to quadratic order about the frequency $\fs (t)$ at which $\phi(f)$ is stationary. The result is:
\be
C(t) = \frac{Q h_{0} e^{-i \pi/4}}{4 i \sqrt{\pi} \f0}
\frac{1}{\fs^{\frac{7}{6}} {S_{h}(\fs)}} e^{i \phi(\fs)} 
\int_{\fl}^{\fu} {e^{- \frac{(f-\f0)^{2}}{2 \ssg^2}}} {e^{i \frac{(f-\fs)^{2}}{2 \sf^2}}} \,df \,,
\label{corr_sg}
\ee
where
\begin{align}
\fs (t)= \f0 \left(\frac{t}{\t0}\right)^{-3/8} \,,
\label{SP}
\end{align}
and
\be
\label{Eq_sigma_f}
\sf(t) = \sqrt{\frac{3 \f0}{16 \pi \t0}} \left(\frac{\fs (t)}{\f0} \right)^{11/6} \,,~~~ \ssg = \frac{\sqrt{2} \f0}{Q} \,.
\ee
Thus, the correlation integrand is expressed as a product of a real
Gaussian centered at $\f0$ with standard deviation $\ssg$ and a
complex Gaussian centered at $\fs$ with a standard deviation
$\sf$. Moreover, since $f^{-7/6}$ and ${S_{h}(f)}$ are slowly varying
functions of $f$ compared to the Gaussians, pulling them out of the 
integral and using their values at  $\fs$ inflicts negligible error on
$C(t)$. Performing the integral in the complex plane and using Cauchy's theorem we arrive at:
\be
\label{SNR}
\rho(t) \equiv |C(t)| = \h \frac{h_{0} \Sigma}{\fs^{7/6}(t) S_{h}(\fs (t)) ({1+ \Sigma^{4}})^{1/4}} 
 e^{- \frac{Q^2}{4} \left (1 - \frac{\fs}{\f0} \right)^2 \frac{\Sigma^4}{1+ \Sigma^{4}}} \,,
\ee
where $\Sigma = \sf / \ssg$, the ratio of the standard
deviations. This was one of the results in Ref.~\cite{Canton:2013joa}. 
\par
Here we are interested in the maximum value of the time-series, $\rho (t)$, and the time at which that maximum occurs. To 
facilitate the calculations and gain insight into the behavior of this
function, we write $\Sigma$ as
\be
\Sigma (t) = \frac{Q}{Q_0} \left (\frac{\fs (t)}{\f0} \right )^{11/6} \,,
\ee
where $Q_0$ is a critical value of the quality factor, and is given by:
\be
Q_0 = \sqrt{\frac{32 \pi \f0 \tau_0}{3}} \,.
\ee
For instance, if the chirp mass is $\M = 1.5~\Msun$ and $\f0 = 60$~Hz, one
finds that $\tau_0 \sim 6$~sec and $Q_0 \sim 110$. 
The critical quality factor $Q_0$ is related to the chirping rate of
the template at $f=f_0$, which we denote by $\dot{f}_0$. 
This rate can be shown to be~\cite{Pai:2000zt}:
\begin{align}
\label{Eq_f0dot}
\dot{f_0} = \frac{3}{8} \frac{f_0}{\tau_0}  \,,
\end{align}
and, hence,
\begin{align}
\label{Eq_Q0_f0}
Q_0 = \frac{2\, \pi \, f_0}{\sqrt{\pi \, \dot{f}_0}} =  2\, \pi \, f_0 \, \tautemp =\sqrt{2} \frac{\omega_0}{\sqrt{\dot{\omega}_0}}  \,,
\end{align}
where $\tautemp=(\pi\, \dot{f}_0)^{-1/2}$ is the characteristic time of the chirping rate of the template at
frequency $f_0$ and $\omega_0 = 2 \pi f_0$. Therefore, $Q_0$ is just the number of cycles corresponding to the characterstic time $\tautemp$.

\subsection{SNRs and  time-lags of sine-Gaussian triggers}
\label{subsec:sgburstsSNRTimelag}

The time-lag $t_m$ and the SNR $\rho_m$ correspond to the maximum of $\rho(t)$, i.e., 
\be
\label{eq:rhomax}
 \max_t \rho(t) = \rho(t_m) = \rho_m \,.
\ee
Unless otherwise mentioned, $\rho(t)$ will always denote the time-series produced by 
a sine-Gaussian glitch $s(t)$ when match-filtered with the unit-norm
inspiral template of Eq.~(\ref{eq:newtonianinspiral}).

We can define an SNR for the sine-Gaussian glitch in a similar manner. We replace $\tilde{h}^*(f)$ in
Eq.~(\ref{crosscorr}) with $\tilde{s}^*(f)$, and normalize as in Eq.~(\ref{eq:normalization}) to obtain the
SNR of the glitch itself, with $s_0$ in Eq. (\ref{eq:sg}) not necessarily equal to unity. The relationship between
the SNR and $s_0$, the amplitude of the glitch, is then
\be
\label{eq:SG_SNR}
\rho_{\mathrm{SG}} = s_0 \sqrt{\int_{\fl}^{\fu} \frac{|{\tilde s}(f)|^2}{S_{h}(f)} \,df}\,.
\ee

The maximization in Eq.~(\ref{eq:rhomax}) is facilitated by going over to the variable $u$ defined by:
\be
u = \frac{\fs}{\f0} = \left (\frac{t}{\tau_0} \right )^{-3/8} \,.
\ee
Assuming that the maximum of $\rho(t)$ occurs when $\fs$ is close to $\f0$ we can write the expression for $\rho$ as:
\be
\rho(u) = \h h_0 \f0^{-7/6} \frac{Q}{Q_0} \left [1 + \left (\frac{Q}{Q_0} \right )^4 \right ]^{-1/4}  \frac{u^{2/3}}{S_h (\f0 u)} e^{- \frac{Q^2}{4} (1 - u)^2} \,.
\label{SNRu}
\ee
Here we have taken the factor $\Sigma^4 / (1+ \Sigma^4)$ to be unity
in the exponential in  Eq. (\ref{SNR}). This is a good approximation as 
long as $\Sigma \gsim 2$. That, in turn, restricts our results'
validity to glitches with $Q \gsim 5$. For glitches that affect the
CBC searches, $Q$ is found to obey that condition.
\par
\begin{figure}[h!]
\begin{center}
 \includegraphics[width=3.5in]{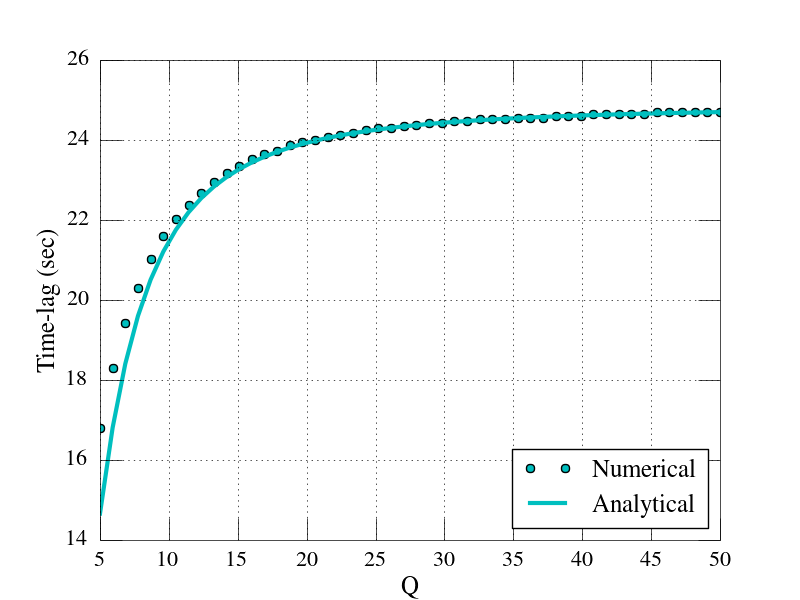}
\end{center}
\caption{The figure shows the time-lags in seconds for $Q$ ranging
  from 5 to 50, $\M = 1.2 \Msun$ and $f_0 = 40$ Hz. The dots are
  obtained numerically by computing $|C(t)|$ whereas the continuous
  curve is the plot of the analytical expression in Eq.~(\ref{eq:tmaxsgburst}).
} 
\label{fig:timedelayvsQMc1p5}
\end{figure}

The maximum of $\rho(t)$ is computed to the linear order in the logarithm of the PSD. Let $\varsigma$ be the negative of the logarithmic derivative of the PSD evaluated at $\f0$, namely,  
\be
\varsigma = - \f0 \frac{S_h' (\f0)}{S_h (\f0)} \,,
\label{ln_der_PSD}
\ee
where the prime denotes the derivative with respect to $f$. Generally,
for the early aLIGO PSD, $\varsigma$ goes from $\sim 4$ to $\sim 0$ as
the frequency varies from 30 Hz to 200 Hz. The value of $u$ that
maximizes $\rho$ is obtained by differentiating Eq. (\ref{SNRu}) and
setting it to zero. Doing so yields a quadratic equation in $u$ from
which the time $t_m$ at which $\rho$ is maximized is obtained. Solving the quadratic we have,
\be
\label{eq:tmaxsgburst}
t_m = \left[\h \left (1 + \frac{2 \varsigma}{Q^2} + \sqrt{\left(1 + \frac{2 \varsigma}{Q^2} \right)^2 + \frac{16}{3 Q^2}} \right) \right]^{-8/3} \tau_0 \,.
\ee
When $Q \gsim 10$, we can simplify the above expression to the leading
order in $1/Q^2$, which yields:
\be
t_m = \left [1 - \frac{16}{3 Q^2} \left(\varsigma + \frac{2}{3} \right) \right] \tau_0 \,.
\ee
The above expressions then yield the time-lag of the trigger. Figure \ref{fig:timedelayvsQMc1p5} shows the time-lags for various $Q$ for $\M = 1.2 \Msun$ and $f_0 = 40$ Hz.

\par

\begin{figure}[ht!]
\begin{center}
\includegraphics[width=3.5in]{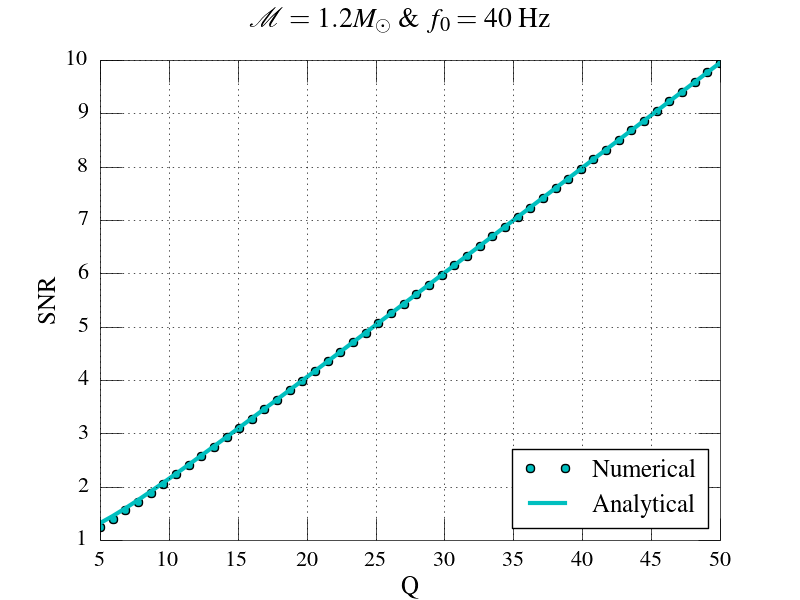}
 \includegraphics[width=3.5in]{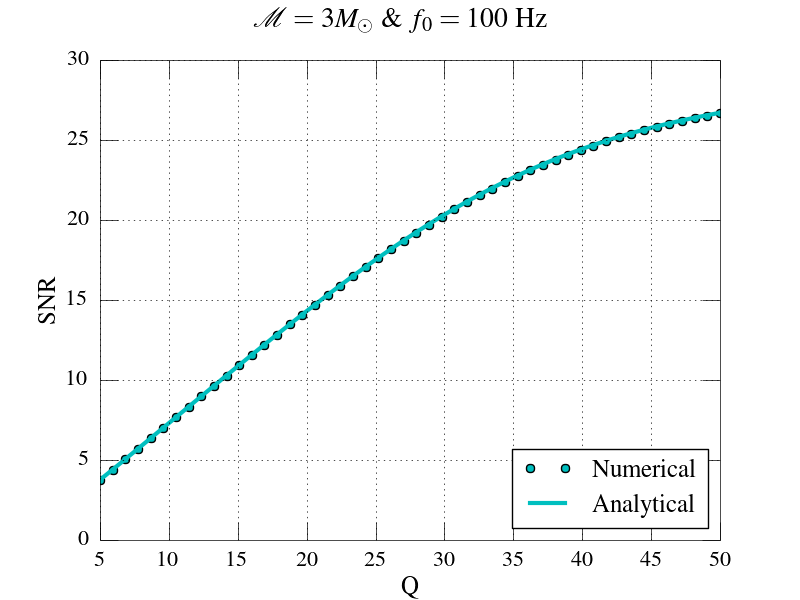}
\end{center}
\caption{The figures show the SNRs of various triggers produced by
  sine-Gaussians with $Q$ values ranging from 5 to 50, when
  match-filtered with a template with $\M = 1.2 \Msun$ (\& $f_0 = 40$
  Hz), as shown in the left panel, and a template with $\M = 3 \Msun$
  (\& $f_0 = 100$ Hz), as shown in the right panel. 
The dots are obtained numerically whereas the continuous curve is the
analytical prediction. Here, we have normalized the SNRs such that 
a sine-Gaussian with $Q=50$ and $f_0=40$~Hz produces a trigger with
SNR of 10 when match-filtered with a binary inspiral template of 
chirp mass $1.2\M_{\odot}$.} 
\label{fig:timedelayvsQMc3p0}
\end{figure}

From $t_m$, the SNR of the trigger up to the order $1/Q^2$ is easily computed, and is given by:
\be
\label{eq:rhomaxsgburst}
\rho_m = \h h_0 \frac{1}{\f0^{7/6} S_h (\f0)} \left( \frac{Q}{Q_0}
\right) \left [1 + \left (\frac{Q}{Q_0} \right )^4 \right ]^{-1/4}
\left [1 + \frac{\left( \varsigma + \frac{2}{3} \right)^2}{Q^2} \right] \,.
\ee
In Fig. \ref{fig:timedelayvsQMc3p0},  the SNR is plotted for $\M = 1.2 \Msun,~f_0 = 40$ Hz and $\M = 3 \Msun,~f_0 = 100$ Hz, while $Q$ ranges from 5 to 50.
In Fig.~\ref{fig:timedelayvsQMc1p5} and \ref{fig:timedelayvsQMc3p0},
we see that our analytical approximations for time-lag and SNR match
very well with numerical results, which are obtained by computing
Eq.~(\ref{crosscorr}), and finding its maximum magnitude and the time
at which it is attained.

\begin{figure}[h!]
\begin{center}
\includegraphics[width=3.5in]{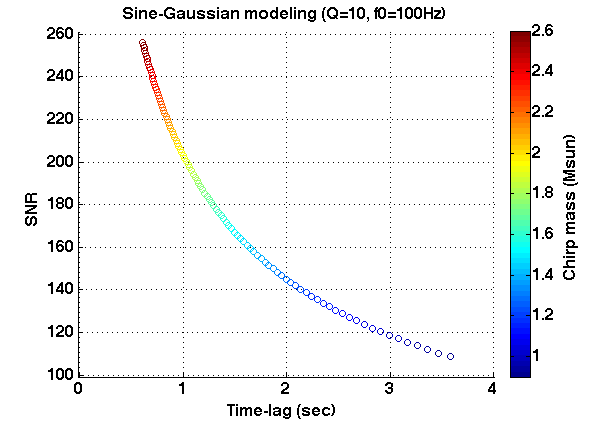}
\includegraphics[width=3.3in]{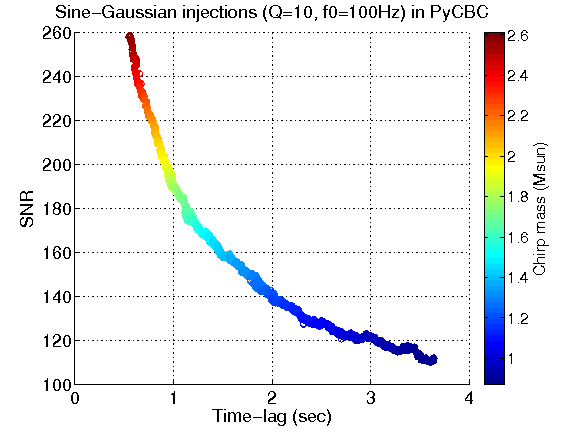}
\end{center}
\caption{The left-hand panel shows the SNR of triggers, $\rho_m$,
  versus time-lag, $t_m$, as predicted by
  Eqs.~(\ref{eq:rhomaxsgburst}) and (\ref{eq:tmaxsgburst}), and obtained
  by our analysis of the effect of a sine-Gaussian glitch with $Q=10$ and $f_0 = 100$ Hz on 
  non-spinning binary inspiral templates. The chirp-mass of the templates is shown in
  color. We have used early aLIGO PSD to generate this plot. The
  right-hand panel shows the SNR versus time-lag of 
  triggers produced by a sine-Gaussian glitch with the same parameters injected in simulated early aLIGO 
  data and recovered by a CBC search pipeline~\cite{Canton:2014ena} using 3.5PN TaylorF2 templates
 \cite{Sathyaprakash:1991mt}. Note the excellent agreement between the two plots.
} 
\label{fig:simSNRVsTimelag4sgburst}
\end{figure}

We observe that for relatively lower values of the chirp mass (e.g.,
$\M=1.2M_{\odot}$) the SNR grows linearly with $Q$, for all values of
$Q$. However, for higher chirp masses, like $\M=3M_{\odot}$, the SNR 
first grows linearly with $Q$, as long as $Q$ is small. This is
because when $Q \ll Q_0$ the nonlinear term $(1  + (Q/Q_0)^4)^{-1/4}$
is nearly constant, and approximately equal to unity; there SNR
$\propto Q$. The critical quality factor $Q_0$ is $\simeq 40$ for the parameters 
chosen in the right panel of Fig.~\ref{fig:timedelayvsQMc3p0}. We observe that when $Q$ increases beyond $\sim 20$, the SNR departs from the straight line. 
Since the fourth power of $Q/Q_0$ is involved, $Q$ does not have to be much less than $Q_0$ for the SNR to increase 
linearly; in fact, $Q \lsim Q_0/2$ is sufficient for the term $(1  + (Q/Q_0)^4)^{-1/4} \simeq 1$ and the SNR to depend linearly on $Q$.  

The SNR versus time-lag, as predicted by Eqs.~(\ref{eq:tmaxsgburst})
  and (\ref{eq:rhomaxsgburst}),  obtained
  from our analysis of the effect of a sine-Gaussian glitch on 
  non-spinning templates is plotted in the left panel of 
  Fig.~\ref{fig:simSNRVsTimelag4sgburst}. This prediction from
  our analytic modeling is in excellent agreement with the behavior
  found  by running the PyCBC search pipeline and is shown in the 
  right panel in that figure.

\begin{figure}[ht!]
\begin{center}
\includegraphics[width=3.5in]{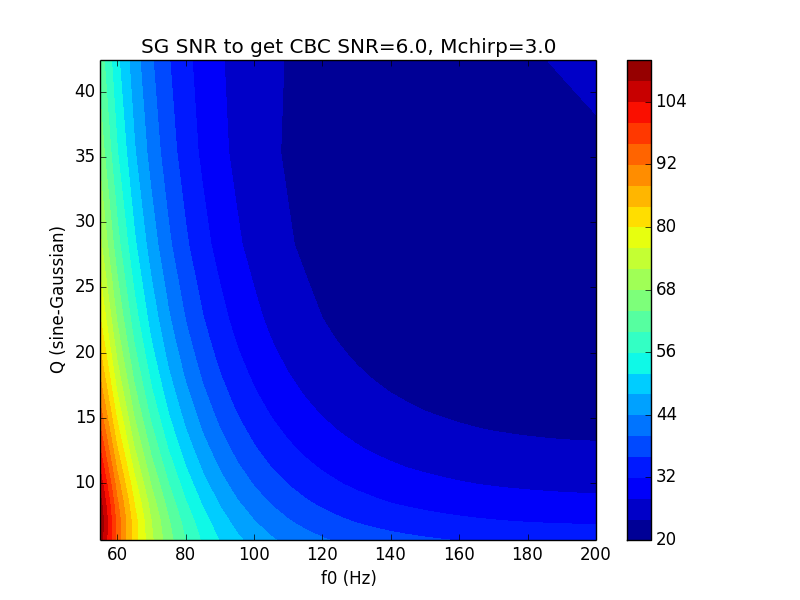}
\end{center}
\caption{The contour plot shows the sine-Gaussian parameters $Q$ and $f_0$ and the colour shows the 
SNR of the sine-Gaussian required to produce a CBC trigger of SNR = 6 when a template with chirp mass $\M = 3 \Msun$ is used. The sine-Gaussian SNR is defined by Eq.~(\ref{eq:SG_SNR}).} 
\label{fig:snr_contourMc3p0}
\end{figure}

Another question of interest is how strong must a sine-Gaussian
glitch be for it to trigger a binary inspiral template with a fixed SNR.
Figure~\ref{fig:snr_contourMc3p0} is a contour plot 
  that shows what sine-Gaussian SNR is required to create a trigger
  with an SNR of 6 when match-filtered with a CBC template of chirp mass $\M = 3 \Msun$.
  The sine-Gaussian's parameters are shown with $Q$ on the vertical axis and $f_0$ on the horizontal axis. The 
upper right-hand corner of the plot, which corresponds to high values
of both $Q$ and $f_0$, can produce a trigger of 
SNR=6 for a relatively low amplitude of the glitch, say, around
20. But as one comes down the diagonal to relatively lower values of the
glitch parameters, e.g., $Q \sim 10$ and $f_0$ below $100$ Hz, one
finds that stronger glitches are needed to produce a trigger of 
the same SNR. The figure outlines the trend of glitch SNRs required to produce a trigger SNR of 6.  

\section{Chirping sine-Gaussian glitches and their effect on Inspiral Templates}
\label{sec:csgbursts}

\begin{figure}[h!]
\begin{center}
\includegraphics[width=3.5in]{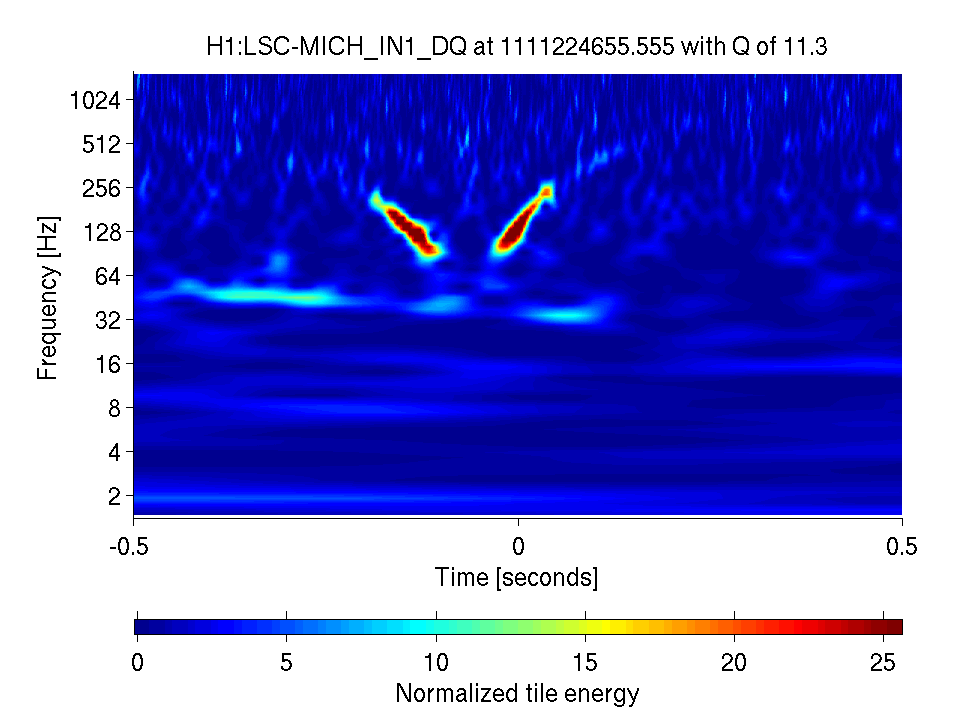}
\end{center}
\caption{The spectrogram shows a noise transient
  with two lobes that lasted approximately from -0.2 sec to 0.1 sec in
  LIGO-Hanford during its commissioning. It was detected by Omicron
  using a sine-Gaussian filter with $f_0 = 121.5$Hz and $Q=11.3$. In
  each lobe, one can find that the frequency of the transient is
  either decreasing (as in the left lobe) or increasing (as in the
  right lobe). The above glitch can be modeled as a
  combination of two chirping sine-Gaussians, with two different
  $\kappa$ values (see Eq.~(\ref{eq:chirpingSG})). 
}
\label{fig:realPipeSNRVsTimelag4Csgburst}
\end{figure}

There exists another class of glitches that occur in the GW detector
data that can be modelled as having a Gaussian envelope, like the sine-Gaussian glitches,
but a rising frequency, unlike the sine-Gaussian glitches. Such glitches are known
as `chirping sine-Gaussians'. Figure \ref{fig:realPipeSNRVsTimelag4Csgburst} shows the spectrogram of 
a noise transient with two lobes that lasted approximately from -0.2 sec to 0.1 sec in LIGO-Hanford during 
its aLIGO commissioning. It was detected by Omicron using a sine-Gaussian filter with $f_0 = 121.5$ Hz and $Q=11.3$. 
In each lobe, one can find that the frequency of the transient is
either decreasing (as in the left lobe) or increasing (as in the right
lobe). These lobes can be modeled as chirping
sine-Gaussian. Just as in the case of the sine-Gaussian glitch, a
strong chirping sine-Gaussian glitch also produces a trail of triggers
when match-filtered with CBC templates. 
It is to model these type of transients that we study the chirping sine-Gaussians.

\begin{figure}[ht!]
\begin{center}
$\begin{array}{cc}
 \includegraphics[width=3.5in]{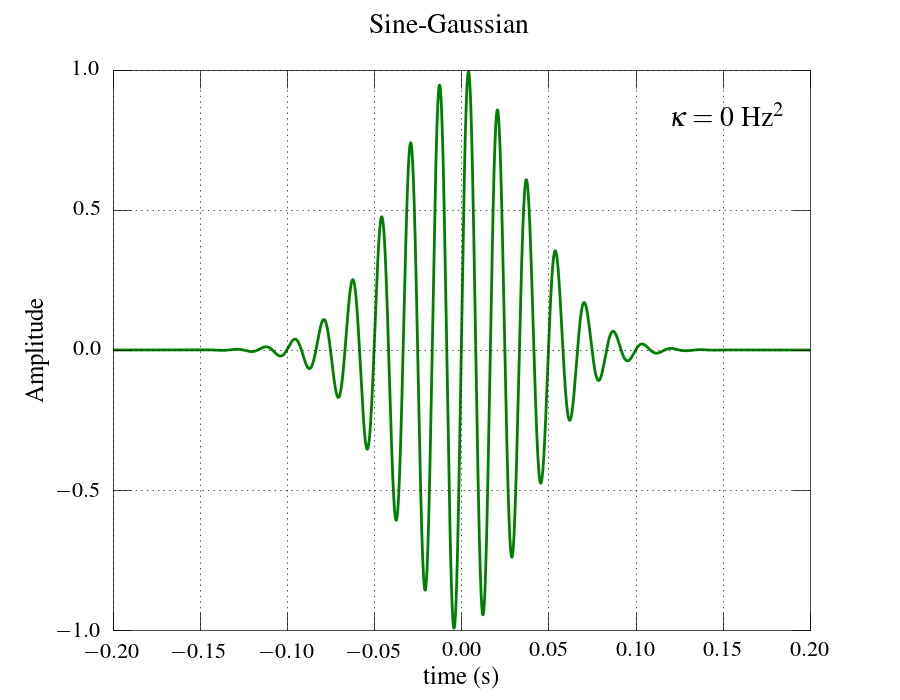}&
 \includegraphics[width=3.5in]{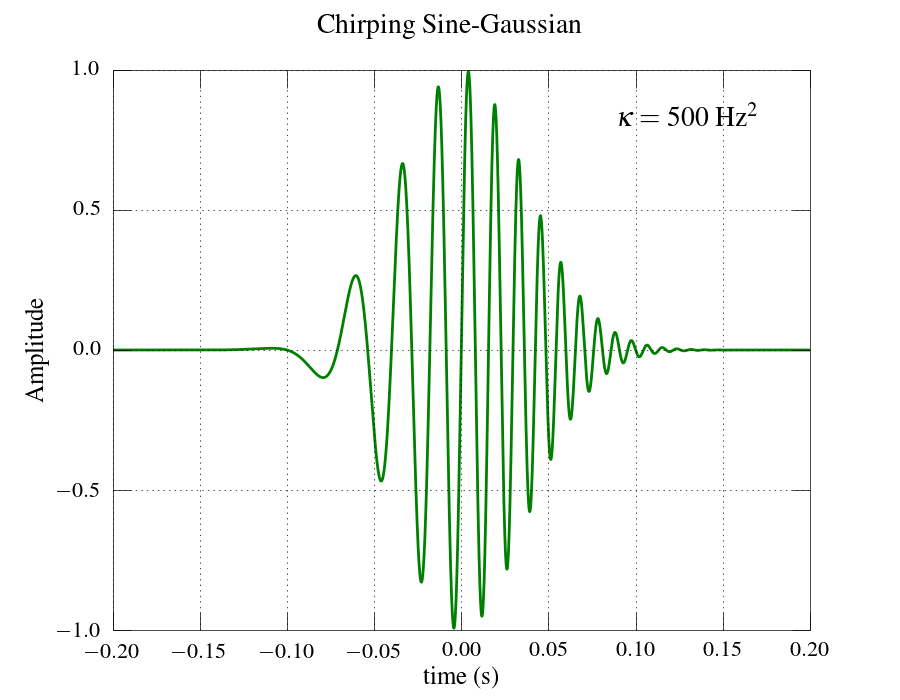}
\end{array}$
\end{center}
\caption{Left panel: unit amplitude time-domain sine-Gaussian with central frequency $f_0=60$ Hz, quality factor $Q=20$ and chirping rate $\kappa = 0$ Hz$^2$.
Right panel: unit amplitude time-domain chirping sine-Gaussian with central frequency $f_0=60$ Hz, quality factor $Q=20$ and chirping 
rate $\kappa=500$ Hz$^{2}$.} 
\label{Fig_SG_CSG_timedomain}
\end{figure}

A chirping sine-Gaussian can be characterized by three parameters:
the central frequency $f_0$, the quality factor $Q$ and chirping rate 
$\kappa$. The expression for a time-domain chirping 
sine-Gaussian reads as follows:
\begin{align}
\label{eq:chirpingSG}
s(t)= s_0  e^{-\frac{t^{2}}{ \tau^{2}}} \sin ( 2 \pi \f0 t + \pi \kappa t^2)\,,
\end{align}
where $s_0$ is the amplitude (which is set to unity, unless otherwise
specified). Figure~\ref{Fig_SG_CSG_timedomain} shows how a chirping sine-Gaussian
glitch looks as compared to a non-chirping one in the time-domain.
Note that the chirping sine-Gaussian reduces to the usual sine-Gaussian, studied in the last section, when 
$\kappa =0$. 
\par 

In the frequency domain the chirping sine-Gaussian for $f > 0$ can be written as,
\be
{\tilde s}(f) = \frac{1}{4 i \sqrt {\pi}} \frac{Q}{\f0} \frac{1}{\bigl(1+ \frac{Q^4}{Q_{\kappa}^4 }\bigr)^{1/4}}
\exp \left [{- \frac{(f-\f0)^{2} Q^{2}}{4 \f0^{2}\,\bigl(1-\frac{i Q^2}{Q_{\kappa}^2}\bigr)}} 
 + \frac{i}{2} \tan^{-1}\biggl(\frac{Q^2}{Q_{\kappa}^2}\biggr) \right ] \,,
\ee
where,
\be
\label{Eq_Qs}
Q_{\kappa}  = \frac{2\, \pi\, f_0}{\sqrt{\pi \, \kappa}} \equiv 2 \pi f_0 \tau_{\kappa} \,,
\ee
and
$\tau_{\kappa} = (\pi \kappa)^{-1/2}$ is the characteristic time over
which the chirping sine-Gaussian changes its frequency.
In the following, we investigate how chirping sine-Gaussian glitches affect the matched filter searches of inspiralling compact binaries.

\subsection{The correlation of a chirping sine-Gaussian and an  inspiral template}
\label{subsec:corrChirpSG}

In this section, we compute the correlation between a chirping
sine-Gaussian and a (Newtonian) inspiral template.
We begin by writing down the correlation integral
\be
\label{Eq_CorrInt}
C(t) = \int_{\fl}^{\fu} \frac{{\tilde s}(f) {\tilde h}^{*}(f)}{S_{h}(f)} e^{2 \pi i f t} \,df 
     \equiv  B\,\int_{\fl}^{\fu} \frac{\exp \biggl(- \frac{(f-\f0)^{2} Q^{2}}{4 \f0^{2} \bigl(1-\frac{i Q^2}{Q_{\kappa}^2}\bigr)} \biggr) \, e^{i \phi(f)}}{f^{7/6} S_{h}(f)}~df \,,
\ee
where 
\begin{align}
B= \frac{h_0}{4 i \sqrt {\pi}} \frac{Q}{\f0} \, \frac{e^{-i\pi/4}}{\bigl(1+ \frac{ Q^4}{Q_{\kappa}^4 }\bigr)^{1/4}} \exp\biggl(\frac{i}{2} \tan^{-1}\biggl(\frac{Q^2}{Q_{\kappa}^2}\biggr)\biggr) \,,
\end{align}
and $\phi(f)$ is given by Eq.~(\ref{phase}). Following analogous arguments as those presented for Eq.~(\ref{corr_sg}) in Sec. \ref{Sec_SG_match}, we can rewrite the correlation integral as
\be
\label{Eq_CorrInte}
C(t) =  \frac{B\, e^{i \phi(f_s)}}{f_s^{7/6} S_h(f_s)} 
\int_{\fl}^{\fu} e^{-\frac{(f-\f0)^{2}}{2 \sigma_r^2}}  e^{-i\frac{(f-\f0)^{2}}{2 \sigma_i^2}} e^{i\frac{(f- f_s)^{2}}{2 \sigma_{\phi}^2}} ~df\,,
\ee
where
\be
\sigma_r =  \frac{\sqrt{2} f_0}{Q} \sqrt{1 + \frac{Q^4}{Q_{\kappa}^4}},~~~\sigma_i =  \sqrt{\frac{8 \pi}{\kappa}} \frac{f_0^2}{Q^2}  \sqrt{1 + \frac{Q^4}{Q_{\kappa}^4}} \,.
\ee
Note that the correlation integral now contains a product of three
Gaussians, where one is real and two are complex. The first and second Gaussians  
are centered at $f_0$ with standard deviations $\sigma_r$  and $\sigma_i$, respectively, while the third Gaussian is centered at $f_s$ with standard deviation 
$\sigma_{\phi}$.
It will be useful to define following dimensionless quantities:
\be
\Delta = \frac{f_0 - f_s(t)}{\sigma_{\phi}} \,,~~~~ \Sigma_r = \frac{\sigma_{\phi}}{\sigma_r} \,, ~~~~\Sigma_i = \frac{\sigma_i}{\sigma_r} \,,~~~~ \frac{1}{\Lambda^2} =  \frac{1}{\Sigma_r^2} -  \frac{1}{\Sigma_i^2} \,.   
\ee
Note that $\Sigma_r$, $\Lambda$ and $\Delta$ are
functions of time through $\sigma_{\phi}$ and $\Lambda = \Sigma$
when $\kappa=0$. (One also finds $\Sigma_r \rightarrow \Sigma$ and $\Sigma_i \rightarrow \infty$ as $\kappa \rightarrow 0$.)
Changing the variable of integration
from $f$ to $x = (f-f_0)/\sigma_{\phi}$ and completing the square in
the integrand, the correlation integral in Eq.~(\ref{Eq_CorrInte}) is expressed as
\be
C(t) = \frac{B \,  \sigma_r \,  e^{i \phi(f_s)} \, e^{\frac{i\Delta^2}{2}} \, e^{- \frac{\Delta^2 \Lambda^2}{2 \Sigma_r^2 (\Lambda^2 -i)}} }{f_s^{7/6} \, S_h(f_s)}
\int_{x_{\rm lower}}^{x_{\rm upper}}  dx ~ \exp \biggl(-\frac{1}{2} \biggl(\frac{\sqrt{\Lambda^2 - i}}{\Lambda} x- \frac{i \Delta \Lambda}{\Sigma_r \sqrt{\Lambda^2 -i}} \biggr)^2\biggr)\,.
\ee
We compute the above integral in the complex plane by using Cauchy's
theorem and obtain
\be
\label{SGC_SNR}
\rho(t) \equiv |C(t)| =  \frac{1}{2} \biggr(1 +  \frac{Q^4}{Q_{\kappa}^4 }\biggr)^{1/4}
\frac{h_0}{f_s(t)^{7/6} S_h(f_s(t))} 
\frac{|\Lambda|}{(1+ \Lambda^4)^{1/4}} e^{- \frac{\Delta^2 \Lambda^4}{2 \Sigma_r^2 (1+ \Lambda^4)}} \,,
\ee
where $\Lambda$ is given in terms of $Q$, $Q_0$ and $Q_{\kappa}$ as follows,
\begin{eqnarray}
\label{Eq_Lambda}
\Lambda = \frac{Q}{Q_0}\, \biggl(\frac{f_s}{f_0}\biggr)^{11/6}
\, \biggr[1+ \frac{Q^4}{Q_{\kappa}^4}\, \bigg(1 - \frac{Q_{\kappa}^2}{Q_0^2}\, \biggl(\frac{f_s}{f_0}\biggr)^{11/3}\bigg)\biggr]^{-1/2}\,.
\end{eqnarray}
It is not very difficult to see that the above equation for $\rho(t)$ reduces to Eq.~(\ref{SNR}) when $\kappa =0$. 
\par

In what follows, we explore physically motivated values of $\kappa$,
which is a free parameter, that give large values of $\rho(t)$. 
We observe that the correlation integral (\ref{Eq_CorrInte}) can be maximized when $f_0 \sim f_s$ and $\sigma_i = \sigma_{\phi}$. 
This leads to the following expression for $\kappa$ (for large $Q$),
\begin{align}
\kappa = \dot{f}_0 = \frac{3}{8}\, \frac{f_0}{\tau_0}  \,.
\end{align}
This indicates that $\rho(t)$ will be larger when the chirping rate of glitch, namely $\kappa$, matches with 
the chirping rate of template $\dot{f}_0$, which is as expected. Note that in the case of sine-Gaussian glitches,
the maximum of $\rho(t)$ occurs around the time when the central frequency of glitch matches with the chirping frequency of inspiral template.
\par

We have observed that the expression for $\rho(t)$ is in excellent agreement with numerical simulations over a wide range of the parameter values. 
However, the approximation fails for very high values of $Q(\gtrsim50)$ when $\kappa \simeq \dot{f}_0$. 
This is because when $\kappa \sim \dot{f}_0$, we have $\sigma_i \sim \sigma_{\phi}$ and for high values of $Q$, $f_s (t_m) \sim f_0$ 
and so two of the Gaussians in the integrand of Eq.~(\ref{Eq_CorrInte}) cancel each other out and only the Gaussian involving $\sigma_r$ is left. 
Typically, we have found $\sigma_r \sim 10$ Hz and therefore treating the other terms, for instance, the PSD, as constants is not justified 
and so cannot be pulled out of the integral. Therefore, the approximation leads to inaccurate results.
\par

Secondly, we have observed that when $\kappa \simeq \dot{f}_0$,
the SNR of a chirping sine-Gaussian is higher than that of its non-chirping
counterpart, with the same $Q$ and $f_0$ but $\kappa =0$. However, when
$\kappa \gg \dot{f}_0$, the SNR of the chirping sine-Gaussian drops
below that of its non-chirping counterpart.

\subsection{Time-lag and SNR for chirping sine-Gaussian triggers }

\label{subsec:timelagCSG}

In this section, we derive the expressions for the time-lag, $t_m$, and the trigger SNR, $\rho_m = \rho(t_m)$,
for chirping sine-Gaussians. The expression for $\rho(t)$ reads,
\begin{eqnarray}
\label{SGC_SNR_II}
\rho(t) &=&  \frac{1}{2} \biggr(1 +  \frac{Q^4}{Q_{\kappa}^4}\biggr)^{1/4} \frac{h_0}{f_0^{7/6} S_h(f_s(t))}\, \frac{Q}{Q_0}\, 
\biggl( \frac{f_s}{f_0}\biggr)^{2/3} \biggr[1+ \frac{Q^4}{Q_{\kappa}^4} \biggl(1 - \frac{Q_{\kappa}^2}{Q_0^2} \, \biggl(\frac{f_s}{f_0}\biggr)^{11/3}\biggr)\biggr]^{-1/2} \nonumber \\
&&\times \biggl\{ 1 + \frac{Q^4}{Q_0^4} \biggl(\frac{f_s}{f_0} \biggr)^{22/3} 
\biggl[ 1+ \frac{Q^4}{Q_{\kappa}^4} 
\biggl(1 - \frac{Q_{\kappa}^2}{Q_0^2} \, \biggl(\frac{f_s}{f_0}\biggr)^{11/3}\biggr) \biggr]^{-2}\biggr\}^{-1/4}  \nonumber \\
&& \times \exp\biggl\{-\frac{Q^2}{4} \biggl(1 - \frac{f_s}{f_0}\biggr)^2 
\biggl[ 1+ \frac{Q^4}{Q_{\kappa}^4}
 \biggl(1 - \frac{Q_{\kappa}^2}{Q_0^2} \biggl(\frac{f_s}{f_0}\biggr)^{11/3}\biggr) \biggr]^{-2} 
\biggr(1 +  \frac{Q^4}{Q_{\kappa}^4}\biggr)
\biggr\}
\,.
\end{eqnarray}
Note that in the above expression for $\rho(t)$ we have dropped the $1/(1+\Lambda^4)$ term appearing in the exponential. We have numerically 
found that this term does not change the results appreciably for the parameters we consider.
In order to maximize $\rho(t)$, we define a smallness parameter `$\epsilon$' 
\begin{eqnarray}
\label{Eq_eplsion}
\frac{f_s (t)}{f_0} = \biggl(\frac{t}{\tau_0}\biggr)^{-3/8} \equiv 1 + \epsilon\,.
\end{eqnarray}
We can then write $S_h(f_0(1+ \epsilon))$ as
\begin{eqnarray}
\frac{1}{S_h(f_0 (1+\epsilon))} \cong \frac{(1+\epsilon)^{\varsigma}}{S_h(f_0)}\,,
\end{eqnarray}
where $\varsigma$ is defined in Eq. (\ref{ln_der_PSD}) as the negative of the derivative of the logarithm of the PSD. Neglecting the higher order $\epsilon$ terms in the expression for
$\rho$ and recasting it in terms of this smallness parameter, we get

\begin{eqnarray}
\rho(\epsilon) &=&  \rho_0 \, (1+ \epsilon)^{\varsigma + 2/3} \, e^{-\mathcal{A}\epsilon^2} \,,
\end{eqnarray}
where
\begin{eqnarray}
\label{SGC_SNR_III}
\rho_0 &=& \frac{1}{2} \biggr(1 +  \frac{Q^4}{Q_{\kappa}^4}\biggr)^{1/4} \frac{h_0}{f_0^{7/6} S_h(f_0)}\, \frac{Q}{Q_0}\, 
\biggr[1+ \frac{Q^4}{Q_{\kappa}^4}\biggl(1 - \frac{Q_{\kappa}^2}{Q_0^2}\,\biggr)\biggr]^{-1/2} \nonumber \\
&&\times \biggl[1 + \frac{Q^4}{Q_0^4} 
\biggl( 1+ \frac{Q^4}{Q_{\kappa}^4}\biggl(1 - \frac{Q_{\kappa}^2}{Q_0^2}\,\biggr) \biggr)^{-2}\biggr]^{-1/4} \,, \\
\mathcal{A} &=& \frac{Q^2}{4} \biggl[ 1+ \frac{Q^4}{Q_{\kappa}^4}\biggl(1 - \frac{Q_{\kappa}^2}{Q_0^2}\biggr) \biggr]^{-2} 
\biggr(1 +  \frac{Q^4}{Q_{\kappa}^4}\biggr)\,.
\end{eqnarray}
Maximizing with respect to $\epsilon$, as was done in the previous
section for the sine-Gaussian glitch, we find that the expression for time-lag is
\begin{eqnarray}
\label{Eq_timelag_CSG}
t_m = \tau_0\, \biggl[1- \frac{4}{3 \mathcal{A}} \, \biggl( \varsigma + \frac{2}{3} \biggr) \biggr] \,.
\end{eqnarray}
Also, the expression for the SNR is readily obtained as
\begin{eqnarray}
\label{Eq_SNR_CSG}
\rho_m = \rho_0\, \biggl[1 + \frac{1}{4 \mathcal{A}} \, \biggl( \varsigma + \frac{2}{3} \biggr)^2 \biggr] \,.
\end{eqnarray}
The above expressions for time-lag and SNR reduce to those for non-chirping sine-Gaussian
glitches when $\kappa = 0$. 

\begin{figure*}
\begin{center}
\includegraphics[width=3.5in]{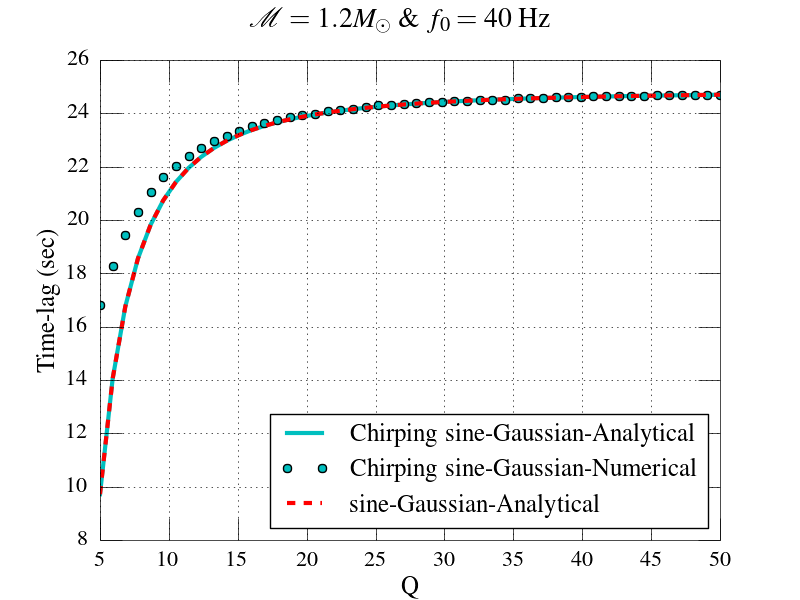}
\includegraphics[width=3.5in]{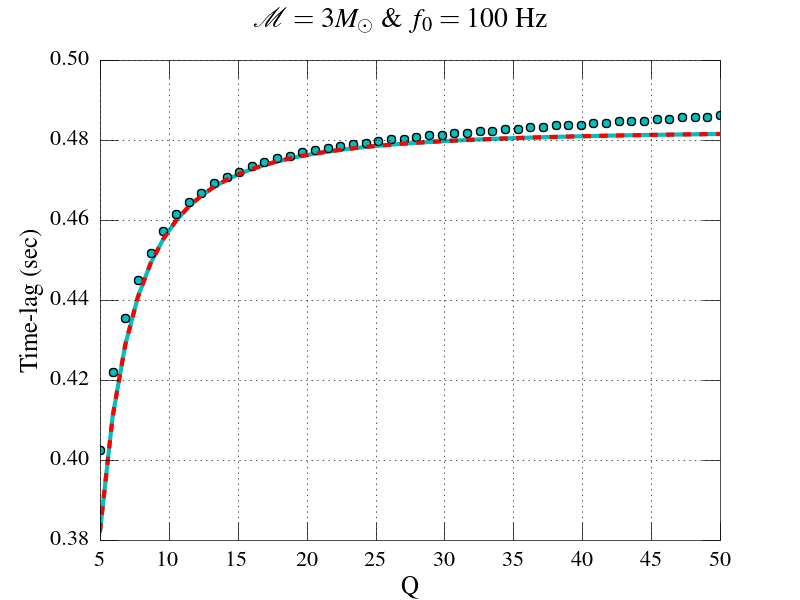} \\
\includegraphics[width=3.5in]{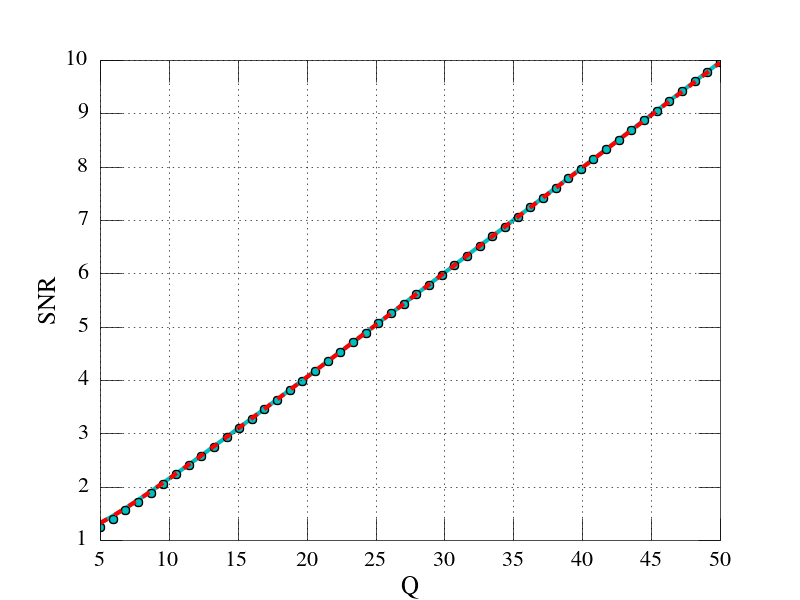}
\includegraphics[width=3.5in]{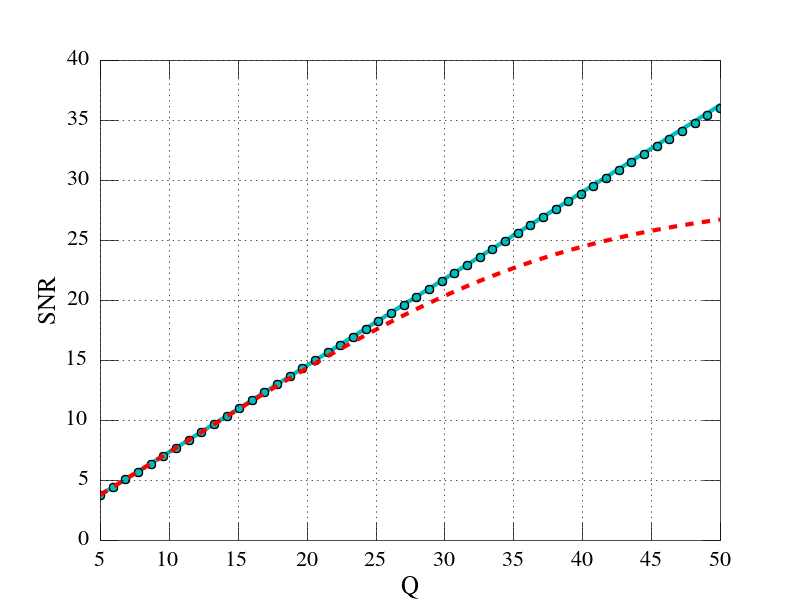} \\
\includegraphics[width=3.5in]{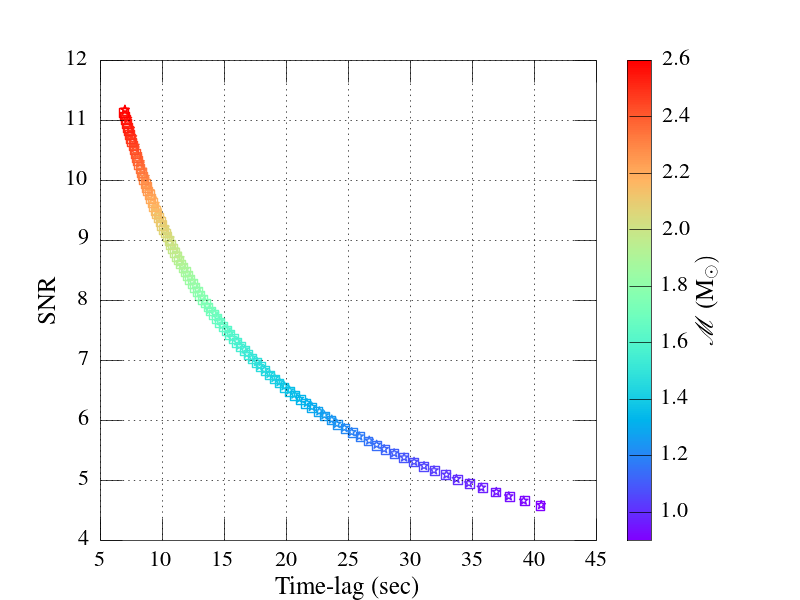}
\includegraphics[width=3.5in]{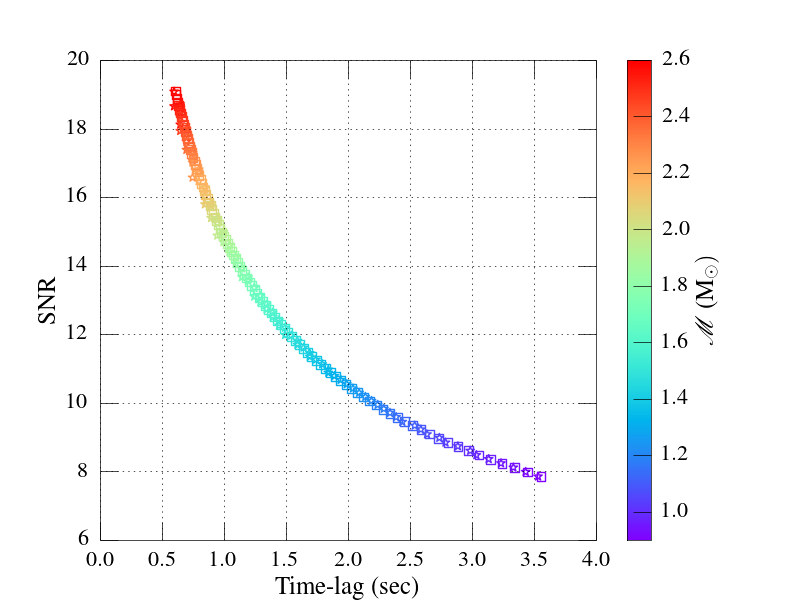}
\end{center}
 \caption{ Plots examining the validity of Eqs.~(\ref{Eq_timelag_CSG}) and (\ref{Eq_SNR_CSG}) for
 trigger time-lag and trigger SNR for $\kappa = \dot{f}_0$. Plots in
 the left panel are for $\mathcal {M} = 1.2M_{\odot}$, $f_0=  100$ Hz \& $\dot{f}_0 = 0.6$ Hz$^2$
 whereas those in the right panel are for $\mathcal {M} = 3M_{\odot}$, $f_0=  100$ Hz \& $\dot{f}_0 = 77.96$ Hz$^2$. 
 Top panel shows the time-lags due to chirping sine-Gaussians as functions of quality factor Q. 
 For $Q>10$, the numerical results (dots) are in good agreement with the analytical ones (solid line). 
 However, for high $Q$ values numerical results deviate from the analytical results in the $\M=3M_{\odot}$ case. This is because the approximation 
for the correlation integral (\ref{Eq_CorrInte}) ceases to be accurate for high $Q$ values.
Moreover, these time-lags are not very different from those for non-chirping
sine-Gaussians (dashed line). 
In the middle panel, the SNR is plotted as a function of Q. 
The numerical results (dots) are in good agreement with the analytic ones (solid line). 
Interestingly, the SNRs for $\M=3M_{\odot}$ are much higher than those for non-chirping sine-Gaussians (red dashed line).
Lower panel depicts SNR versus time-lag plots for a $Q=30$ chirping
sine-Gaussian. The color bar is for chirp mass varying from $0.87M_{\odot}$ to $2.6M_{\odot}$. The numerical 
results are depicted with squares which are in good agreement with the analytical values (stars).
}
\label{Fig_SNR_timelag_Q_b_1}
\end{figure*}

In Fig.~\ref{Fig_SNR_timelag_Q_b_1}, we examine the validity of Eqs.~(\ref{Eq_timelag_CSG}) and (\ref{Eq_SNR_CSG}) for
trigger time-lag and trigger SNR. The left panel has parameters  $\mathcal {M} = 1.2M_{\odot}$, $f_0=  40$ Hz and $\dot{f}_0 = 0.6$ Hz$^2$
 whereas the right panel is for $\mathcal {M} = 3M_{\odot}$, $f_0=
 100$ Hz and $\dot{f}_0 = 77.96$ Hz$^2$. The chirp masses of
 $1.2M_{\odot}$ and $3M_{\odot}$ are taken to characterize a neutron
 star binary and a black-hole-neutron star binary system, respectively.
The top panel shows that the time-lag $t_m$
matches very well with the results from numerical simulations for a
wide range of $Q$ values, between 5 and 50. For comparison, 
we have also plotted the time-lags in the case of the non-chirping 
sine-Gaussians. These plots suggest that the time-lags 
due to chirping and non-chirping sine-Gaussians are not much
different. However, when the template chirp mass is high, the trigger SNRs produced 
by chirping sine-Gaussians can be much higher than those produced by non-chirping sine-Gaussians, as shown in the middle panel; there we 
have plotted the SNRs, $\rho_m$, as functions of $Q$, along 
with their non-chirping counterparts (i.e., $\kappa = 0$). The SNRs resulting from numerical simulations are plotted as 
well to demonstrate the validity of Eq.~(\ref{Eq_SNR_CSG}). 
The lower panel of Fig.~\ref{Fig_SNR_timelag_Q_b_1} depicts the SNR versus time-lag plot.

\begin{figure}[h!]
\begin{center}
\includegraphics[width=3.5in]{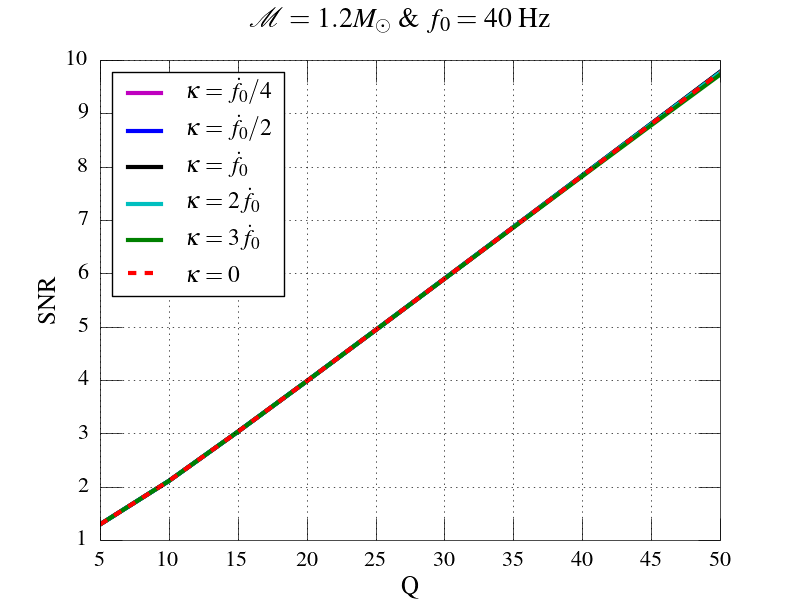}
\includegraphics[width=3.5in]{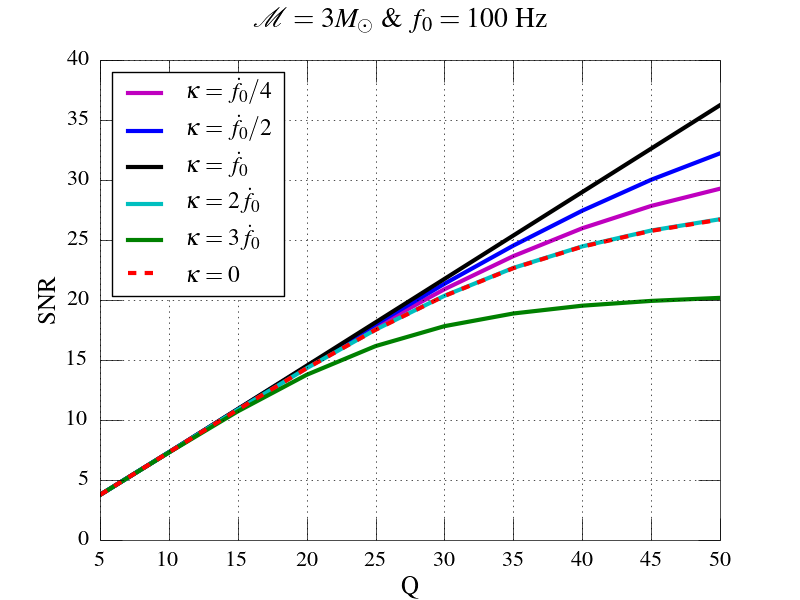}
\end{center}
\caption{The above plots show how the SNR varies with $Q$ for various
  $\kappa$ values, namely, $\kappa=\dot{f}_0/4$, $\dot{f}_0/2$, $\dot{f}_0$, $2\dot{f}_0$, $3\dot{f}_0$.
Left panel is for $\M=1.2M_{\odot}$ \& $f_0=40$ Hz while the right panel is for  $\M=3M_{\odot}$ \& $f_0=100$ Hz. 
In the case of high chirp mass (right panel), as $\kappa$ increases
from zero to larger values the SNR for any $Q$ increases and attains a
maximum for that $Q$ at $\kappa = \dot{f}_0$. Thereafter, the SNR
decreases, and for $\kappa \gg \dot{f}_0$ it drops below the SNR of the
non-chirping sine-Gaussian for the same $Q$. This type of behavior is not seen in the low chirp mass case (left panel) because of the 
very small $\kappa$ values ($\dot{f}_0 = 0.6$ Hz$^{2}$).
}
\label{Fig_SNR_Q_multi_b}
\end{figure}

For comparison, we also study cases when $\kappa$ is quite
different from $\dot{f}_0$. First consider the response of a binary
inspiral template with $\M=3M_{\odot}$ to chirping sine-Gaussians, as shown in Fig.~\ref{Fig_SNR_Q_multi_b}. When $\kappa = \dot{f}_0/2$ the 
behavior is similar to the case $\kappa = \dot{f}_0$ for any given $Q$,
where the chirping sine-Gaussian produces a higher SNR than the
non-chirping one, but not as large as that for the $\kappa =
\dot{f}_0$ case. But when the chirping rate is large, such as in the case of $\kappa = 2 \dot{f}_0$, 
both the chirping as well as the non-chirping sine-Gaussians produce approximately the same SNR. 
Raising $\kappa$ to, say, $3 \dot{f}_0$ reduces the SNR further. In summary, if we start 
from $\kappa = 0$ and increase $\kappa$ steadily, the SNR increases monotonically until $\kappa = \dot{f}_0$, and falls thereafter. When $\kappa \sim 2 \dot{f}_0$, it falls back to the SNR of the non-chirping case and decreases further 
for higher values of $\kappa$; when $\kappa = 3 \dot{f}_0$, the SNR is less than that of the non-chirping case. 
Note that this behavior can not be seen for $\M=1.2M_{\odot}$ case. This is because for 
such a case the value of $\kappa = \dot{f}_0$ is very small ($\sim 0.6$ Hz$^2$) and hence does not show any difference
as compared to non-chirping sine-Gaussian results.

The above analysis can be used to address the problem of designing effective vetoes for these glitches. We do so in the next section. 

\subsection{Post-Newtonian templates}
\begin{figure}[h!]
\begin{center}
\includegraphics[width=3.5in]{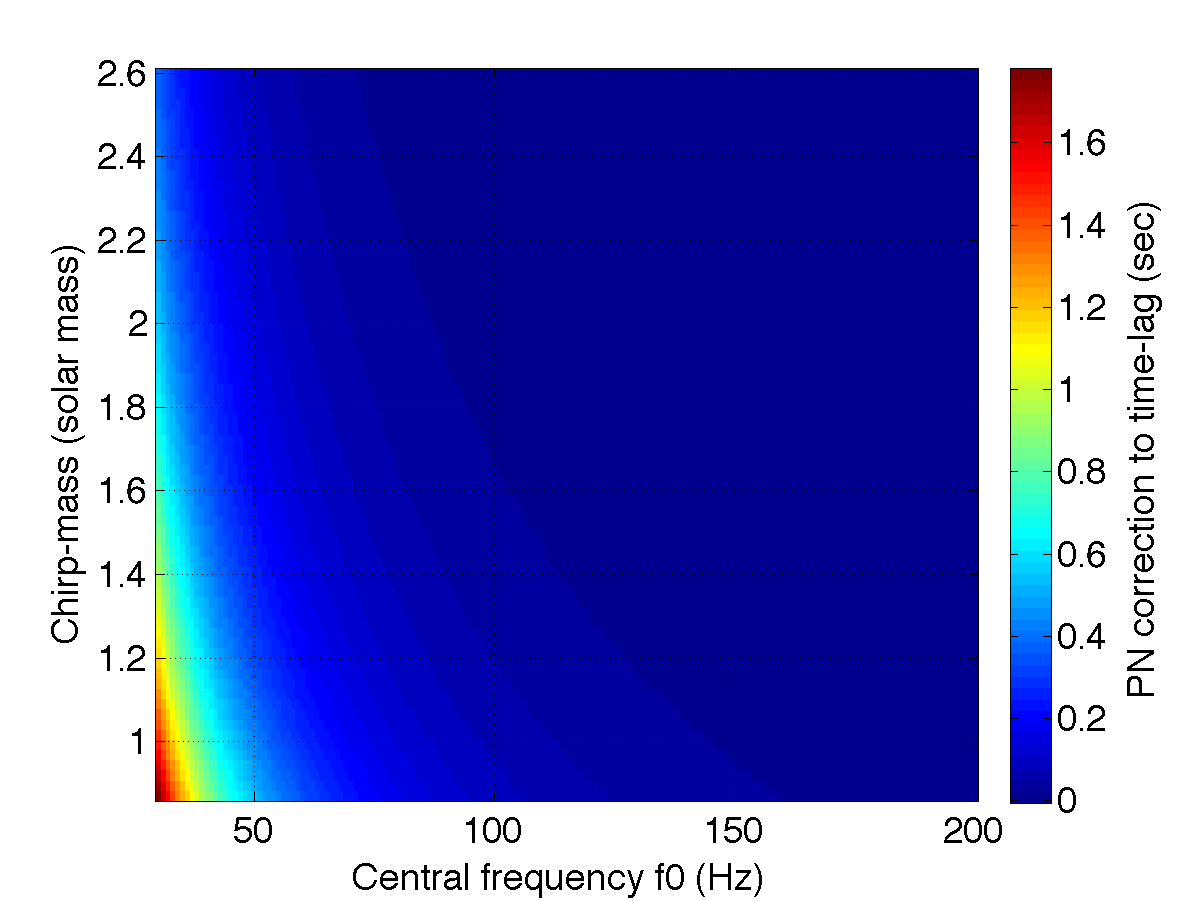}
\includegraphics[width=3.5in]{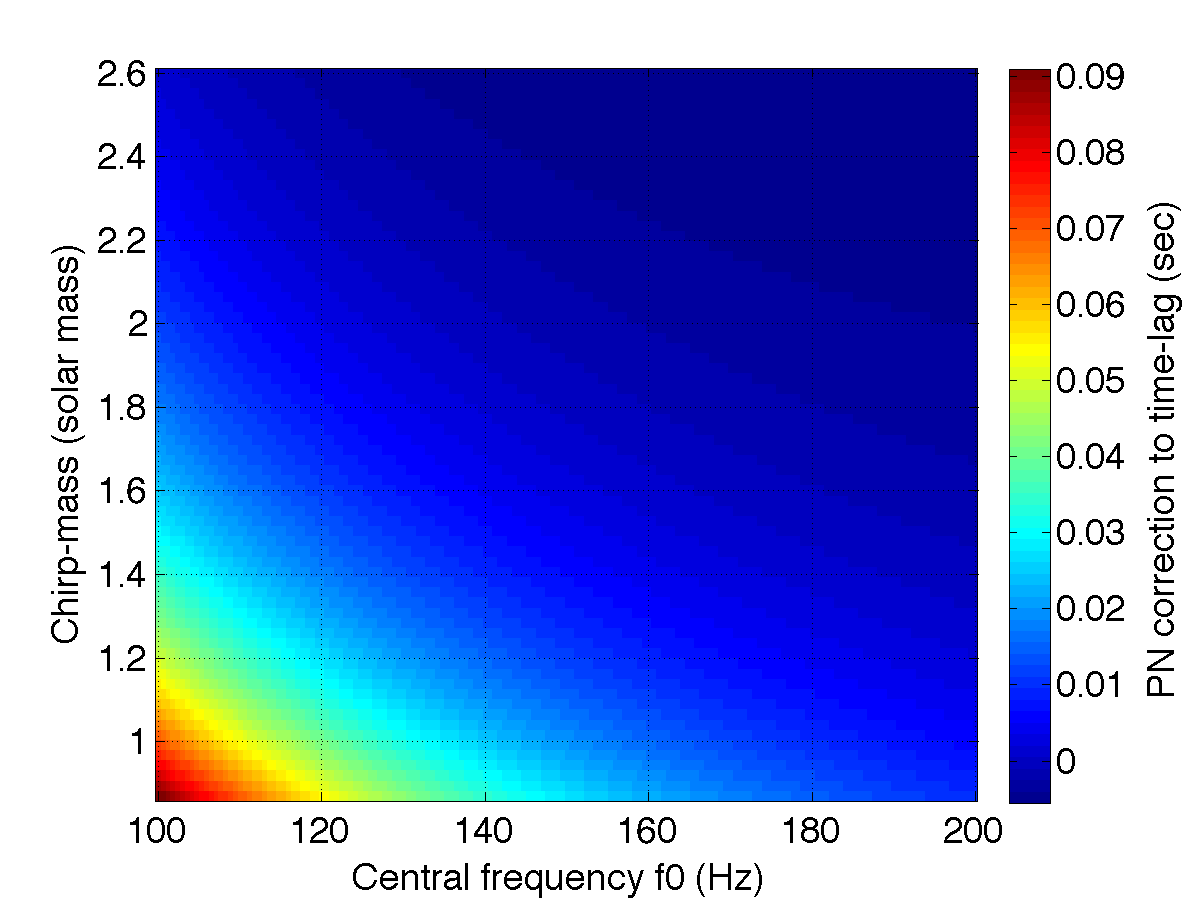}
\end{center}
\caption{
In the left panel we show the post-Newtonian corrections to the time-lag of a trigger that is 
created when a Newtonian template, with chirp-mass $\M \in 
  [0.87,~2.6]M_\odot$, is used to matched-filter a sine-Gaussian 
glitch with quality factor $Q\in [5,50]$ and central frequency 
$f_0 \in [30,200]$Hz. These corrections are obtained after 
maximization over $Q$, in the aforementioned domain, and 
$\eta \in [0.16,0.25]$, and can be as high as 1.8 sec. However, as
shown in the right panel, for $f_0 \gtrsim 100$ Hz, these corrections
are less than 0.1 sec.
} 
\label{fig:pnEffects}
\end{figure}

To keep the interpretation of the effect of glitches simple, we used
inspiral waveforms in the Newtonian approximation, parameterized by
the chirp-mass of the binary. The search pipeline results we show
here, on the other hand, employ more accurate waveforms, namely,
ones based on the TaylorF2 approximant computed up to 3.5PN 
order~\cite{Sathyaprakash:1991mt}. It is therefore important to
examine how the time-lag of a trigger
from a (chirping) sine-Gaussian glitch changes owing to the PN terms,
especially, in the phase of an inspiral template.
The effect of these PN ``corrections'' on the time-lags of a trigger
are depicted in Fig.~\ref{fig:pnEffects}. These triggers are
created when a Newtonian template, with chirp-mass $\M \in 
  [0.87,~2.6]M_\odot$, is used to matched-filter a sine-Gaussian 
glitch with quality factor $Q\in [5,50]$ and central frequency 
$f_0 \in [30,200]$Hz. These corrections are obtained after 
maximization over $Q$, in the aforementioned domain, and 
$\eta \in [0.16,0.25]$, and can be as high as 1.8 sec. However, as
shown in the right panel of that figure, for $f_0 \gtrsim 100$Hz, these corrections
are less than 0.1 sec. The implication
of this finding is that the veto window for tools like H-veto should
be set after accounting for these PN corrections to the Newtonian
time-lags, which we discuss further in the next section.

\section{Translating understanding of glitches to effective vetoes}
\label{sec:vetoes}

\begin{figure}[h!]
\begin{center}
\includegraphics[width=3.5in]{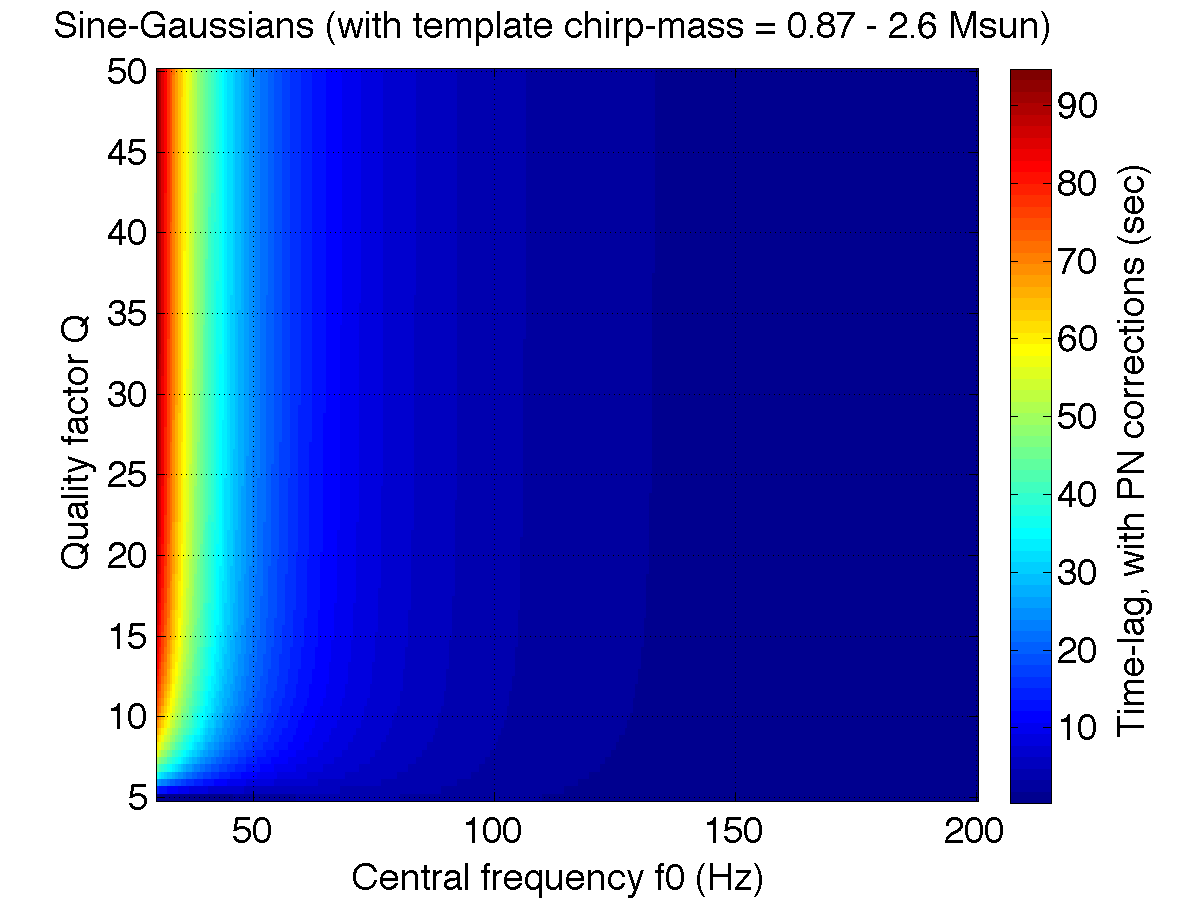}
\includegraphics[width=3.5in]{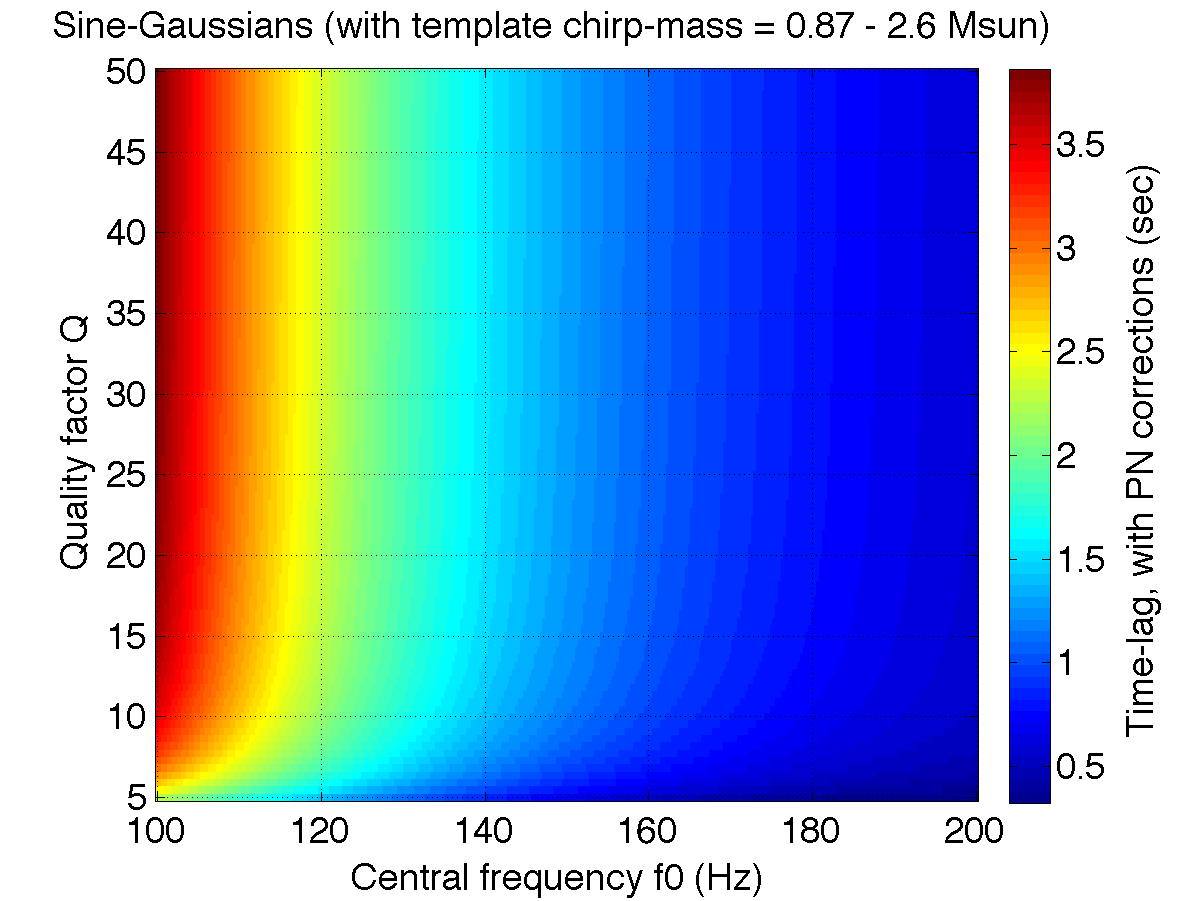}
\caption{The left figure shows the maximum time-lag of binary inspiral triggers produced by various
  types of sine-Gaussian glitches (with $Q\in [5,50]$ and $f_0 \in [30,200]$Hz)
  when a bank of TaylorF2 templates computed up to 3.5PN order, with
  $\M \in [0.87,~2.6]M_\odot$, is used to filter data that contains
  them, one at a time.  What is immediately
  manifest is that glitches with $f_0 \lesssim 50$ Hz 
  produce triggers with the largest time-lags. (Also, the time-lags
  mostly remain unchanged with $Q$, except for very small values of $Q$.)
 Triggers from these glitches can be trickier to veto than those from
 glitches with larger values of $f_0$ because a long time-separation
 makes it challenging to causally associate triggers with
 glitches, and can adversely impact the efficiency of detecting real
 inspiral signals.
In the right figure, same quantity is plotted but for $f_0\in[100, 200]$Hz. The time windows for H-veto-like tools can be as 
small as 4 sec for these glitches.}
\label{fig:sgmaxtimelaginf0qspace}
\end{center}
\end{figure}

We next examine how the understanding developed in the earlier
sections about the effect of certain types of noise transients on CBC templates
can be utilized to improve the sensitivity of CBC searches.~\footnote{A similar question can be 
posed in the context of unmodeled GW signal or astrophysical ``burst'' searches, e.g., 
with multi-detector networks, but that problem has certain additional challenges than the CBC 
searches studied here and will be pursued elsewhere.} 
An important way by which this can be achieved,
in general, is as follows: The time-lag of a CBC trigger arising from a glitch 
can be used to define a veto time ``window'' around the glitch so that the GW 
strain data in that window is labelled as “suspect”. This does not imply that any 
CBC trigger related to that data should not be considered as an event but that more 
checks will need to be performed before it is labelled as one. To wit,
the left panel of Fig. \ref{fig:sgmaxtimelaginf0qspace} 
shows the maximum time-lag of binary inspiral triggers
produced by various  types of sine-Gaussian glitches (with $Q\in
[5,50]$ and $f_0 \in [30,200]$Hz) when a binary inspiral template bank
with $\M \in [0.87,~2.6]M_\odot$ is used to filter data that contains
them, one at a time. These time-lags include PN corrections. 
Their values are mostly unaffected by the quality factor of
the sine-Gaussian glitch, except when $Q \lesssim 10$.
What is immediately manifest is that glitches with $f_0 \lesssim 50$ Hz 
produce triggers with the largest time-lags. Triggers from
these glitches can be trickier to veto than those from glitches with
larger values of $f_0$ because a long time-separation makes it
challenging to causally associate triggers with glitches, and can
adversely impact the efficiency of detecting real inspiral
signals. Therefore, every effort must be made to understand 
what causes these {\em low central-frequency} glitches and reduce their rate of occurrence. 

A {\it prima facie} message of Fig. \ref{fig:sgmaxtimelaginf0qspace}
is that any data quality tool, e.g., “H-veto” \cite{Smith:2011an},
that is used to flag data around glitches based on CBC trigger
time-lags, will need to set time windows that are as 
large as 93 sec for low-mass compact binary searches. While we are 
not implying that these data be 
discarded, such a large time-window nevertheless increases the amount
of data that requires further reanalysis or vetting. 
A better solution is to be more adaptive in selecting the reanalysis
time-windows by utilizing the central frequency $f_0$ of the glitch,
as is reported by data quality tools such as Omicron. As shown in
the left panel of Fig. \ref{fig:sgmaxtimelaginf0qspace} an important finding is that the
maximum time-lag for a bank is independent of $Q$ to a large
extent. Therefore, the primary factor that determines this time-lag is
$f_0$. In fact a large number of glitches are found to have $f_0 \gsim
100$ Hz, for which the maximum time-lag is 4 sec for the template
bank used here (see the right panel of Fig. \ref{fig:sgmaxtimelaginf0qspace}). 
This means that for those types of glitches the
reanalysis time-window can be as short as 4 sec after the
glitch. Glitches with higher $f_0$ will require even shorter windows, thus, reducing the data quality “dead” times \cite{Smith:2011an}.

The utility of our study is demonstrated through observations based on real data
as follows. When the time-series data segment from LIGO's sixth science run (S6)
that contained the glitch shown in
Fig.~\ref{fig:omicronSpect} is filtered with CBC templates with
component masses $\in [1,
3]~M_\odot$, the glitch is found to trigger them with varying time-lags, with a maximum
value of 50 sec. To understand this observation we used Omicron to
find the sine-Gaussian components into which this burst can be
resolved. We found the central frequencies of those components to be
in the range $f_0 \in (35, 50)$ Hz, with various quality factors. We next simulated
the same overlapping sine-Gaussian strains, injected them into 
simulated (Gaussian) early aLIGO noise data and studied their collective effect
on the same bank of CBC templates. The maximum time-lag produced by the glitch
so simulated on this template bank was found to be 50 seconds. This observation is explained
by the fact that the smallest $f_0$ in the real burst is
35 Hz, which when loud enough can trigger templates with time-lags as large as 50 sec,
as can be deduced from
Eq.~(\ref{eq:tmaxsgburst}). Therefore, time-lags estimated by that expression can be
valuable in setting windows for H-veto.

It is important to note that our proposal for applying adaptive windows
for H-veto is computationally cheap. This is because the
computation of time-lags for various values of $Q$, $f_0$, and
$\kappa$, based on our expressions above, is inexpensive.

\begin{figure}[h!]
\begin{center}
\includegraphics[width=3.5in]{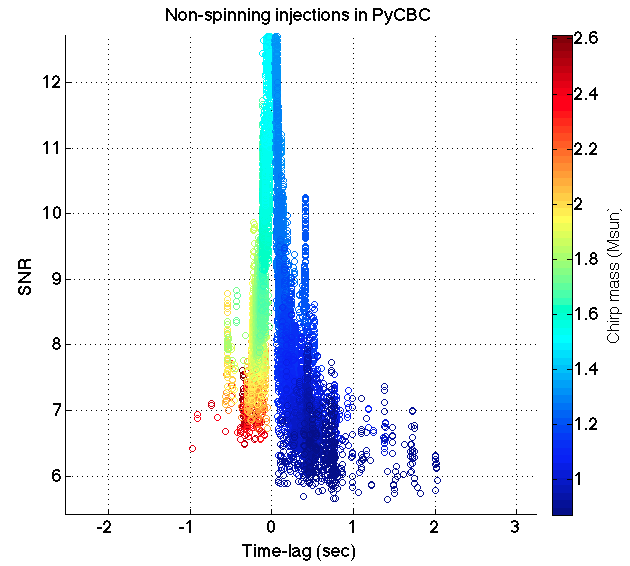}
\caption{The figure shows how the SNR of triggers
  produced by a simulated binary inspiral signal rises and falls as a function of
  the time-lag, with its peak attained at the end time of the signal.}
\label{fig:chirpambiguity}
\end{center}
\end{figure}

\section{Conclusions}
\label{sec:conclusions}

The quest for data analysis search algorithms to distinguish between
glitches and GW signals is complex, in part because 
hunting down the physical cause of glitches is an ongoing
process, owing to the non-stationarity of the data.
An important family of such glitches can be modelled as sine-Gaussians or
chirping sine-Gaussians and is the focus of this paper. In advanced
detectors whose sensitive band begins at low frequencies, i.e., around
10~Hz, CBC templates are of long duration, especially, for low masses. The
glitches studied here can produce triggers long after the glitch has
occurred and, thus, can escape our attention if we search for
glitch-trigger coincidences within short time-windows. Our work serves
to give a mapping from the glitch parameters and template parameters to
the trigger parameters, namely, the SNR and the time-lag, so that such
triggers can be successfully associated with glitches and quarantined
for further vetting. 

One of the main results of this paper is a ready reckoner for
obtaining the inspiral trigger time-lag corresponding to a
sine-Gaussian glitch. For these glitches and a low-mass inspiral
template bank, with $\M \in [0.87,~2.6]M_\odot$, its value, maximized over that bank, can be 
read-off from Fig.~\ref{fig:sgmaxtimelaginf0qspace}, the moment the quality factor and the
central frequency of the glitch are known, e.g., from Omicron.
For chirping sine-Gaussian glitches, Eq.~(\ref{Eq_timelag_CSG}) can be used to find
the time-lags. This, however, requires determination of the glitch
chirp parameter $\kappa$. Therefore, one of the main recommendations
of this work is to develop a tool, or supplement Omicron, to measure
$\kappa$ so that the time-lags for chirping sine-Gaussian glitches can be computed
using the aforementioned formula in an automated fashion.

Another unsurprising result is that low central-frequency glitches
produce the largest trigger time-lags, which makes it difficult to
establish the cause-effect relation between them. Therefore, a second 
recommendation is that every effort be made to determine the
source of these low $f_0$ glitches so that their rate of occurrence is reduced.

Finally, one idea for improving the performance of CBC search
pipelines is as follows. As was shown in Ref.~\cite{Canton:2013joa} and above,
when one filters data containing the types of glitches studied here
with a CBC template bank,
trails of triggers are produced that lie on a hyperbolic curve in a plot of
their SNR versus their time-lag. The shape and duration of this trail
may be different from those of trails produced by CBC signals, as
shown in Fig.~\ref{fig:chirpambiguity}. These distinguishing characteristics may also be used to
construct improved vetoing procedures for isolated glitches in the future.   
\par

\section*{Acknowledgments}

We thank Tito Dal Canton, Josh Smith, Badri Krishnan, Tom Dent, Archana Pai, Peter
Saulson, and Fred Raab for helpful discussions. SB and SVD also thank Bruce Allen and Badri Krishnan for the 
hospitality during their visits to the Albert-Einstein-Institute,
Hannover, where some of the work was carried out. 
This work is supported in part by NSF Grant No. PHY-1206108 and  IUSSTF Award No. 61-2010.
This paper has LIGO document number LIGO-P1600145.


\end{document}